\documentclass[10pt,aps,pre,amsmath,amsfonts,amssymb,floatfix,longbibliography,superscriptaddress,notitlepage, onecolumn]{revtex4-1}
\usepackage{mathrsfs} \usepackage{graphicx,color} \usepackage{bbm} \usepackage{multirow}
\usepackage[normalem]{ulem}
\usepackage{lineno}

\usepackage[colorlinks=true]{hyperref}
\hypersetup{
  colorlinks=true,
  linkcolor=blue,
  filecolor=magenta,
  citecolor=blue,
  urlcolor=blue,
}

\let\a=\alpha  \let\g=\gamma \let\d=\delta

  \let\f=\varphi

   \let\io=\infty

  \def\erf{\text{erf}}

\def\to{\rightarrow}

\newcommand{\beq}{\begin{equation}} \newcommand{\eeq}{\end{equation}}

\renewcommand{\phi}{\varphi}

\def\vr{{\boldsymbol r}}

\newcommand{\YJ}[1]{\textcolor{black}{#1}}

\newcommand{\FZ}[1]{\textcolor{black}{#1}}

\newcommand{\YJJJ}[1]{\textcolor{black}{#1}}
\begin{document}

\title{A stability-reversibility map  unifies elasticity, plasticity, yielding and jamming in hard sphere glasses}

\author{Yuliang Jin}
\email{yuliangjin@itp.ac.cn}
\affiliation{Cybermedia Center, Osaka University, Toyonaka, Osaka 560-0043, Japan}
\affiliation{{\color{black} CAS Key Laboratory for Theoretical Physics, Institute of Theoretical Physics, Chinese Academy of Sciences, Beijing 100190, China}}

\author{Pierfrancesco Urbani}
\affiliation{Institut de physique th\'{e}orique, Universit\'{e} Paris Saclay, CNRS, CEA, F-91191 Gif-sur-Yvette, France}

\author{Francesco Zamponi}
\affiliation{Laboratoire de physique th\'eorique, D\'epartement de physique de l'ENS, \'Ecole normale sup\'erieure, PSL Research University, Sorbonne Universit\'es, CNRS, 75005 Paris, France}

\author{Hajime Yoshino}
\email{yoshino@cmc.osaka-u.ac.jp}
\affiliation{Cybermedia Center, Osaka University, Toyonaka, Osaka 560-0043, Japan}
\affiliation{Graduate School of Science, Osaka University, Toyonaka, Osaka 560-0043, Japan}

\begin{abstract}
  Amorphous solids, such as glasses, have complex responses to deformations, 
  with significant consequences in material design and applications. 
In this respect two intertwined aspects are important: \emph{stability} and \emph{reversibility}. 
It is crucial to understand on the one hand 
how a glass may become unstable due to increased plasticity under shear deformations;
on the other hand,
to what extent the response is reversible, meaning how much a system is able to recover the original configuration once the perturbation is released. 
{\color{black} Here we focus on assemblies of hard spheres as the simplest model of amorphous solids such as colloidal glasses and granular matter. 
We prepare glass states quenched from equilibrium supercooled liquid states, which are obtained by using the swap Monte Carlo algorithm and correspond to a wide range of {\color{black}structural relaxation} time scales.
We exhaustively map out their stability and reversibility} under {\color{black} volume} and shear strains, using extensive numerical simulations.
The region on the {\color{black} volume}-shear strain phase diagram
  where the original glass state remains solid
  is bounded by the shear-yielding
  and the shear-jamming lines which meet at a yielding-jamming crossover point.
  This solid phase can be further divided into two sub-phases: the {\color{black} {\it stable glass phase}} where the system
  deforms purely elastically and is totally reversible, and the {\color{black} {\it marginal glass   phase}
} where it experiences stochastic plastic deformations at mesoscopic scales and  is partially irreversible.
The details of the stability-reversibility map depend strongly on the quality of annealing of the glass. 
This study provides a unified framework for understanding elasticity, plasticity, yielding and jamming in amorphous solids.
\end{abstract}

\maketitle

\section*{Introduction}

Understanding the response of amorphous materials to deformations is a central problem in condensed matter 
both from fundamental and practical viewpoints.
It is not only a way to probe the nature of amorphous solids and their properties, but also crucial to understand a wide range of phenomena from the fracture of metallic glasses to earthquakes and landslides.
Furthermore, it has important applications in material design~\cite{WDS04}.
Although many research efforts have focused on the mechanisms leading to the formation of amorphous 
solids {\color{black}from liquids} ~\cite{angell2000relaxation,Ca09,LN10,CKPUZ17},
an orthogonal
approach is to study these materials deep inside their amorphous phase~\cite{schuh2007mechanical,rodney2011modeling,barrat2011heterogeneities}.
In this work we focus on this second strategy by addressing the problem of understanding the nature of the response of glasses
to {\color{black} volume} and shear strains.

To a first approximation,
glasses are solids much like crystals: they deform essentially elastically for small deformations, but yield under large enough shear strains and start to flow. However, glasses are fundamentally different from crystals, being out-of-equilibrium states of matter. As a consequence, the properties of glasses strongly depend on the details of the preparation protocol~\cite{Ca09}.  {\color{black}
As an example, the yielding of glasses prepared via
a fast quench or very slow annealing is qualitatively different~\cite{ozawa2018random}.} {\color{black} Thus, in sharp contrast to
ordinary states of matter such as gases, liquids and crystals, 
the equations of state (EOS), or the constitutive laws,
of glasses, which characterize their macroscopic  properties,
must depend on the preparation protocol.}
Understanding the mechanical properties of glasses from a unified microscopic point of view
thus emerges as a challenging problem~\cite{bonn2017yield}.

To this aim, a central question is to understand the degree of \emph{stability} of a glass,
i.e. to what extent it can resist to deformations.
{\color{black} In isotropic materials such as glasses,
  it is sufficient to consider two types
  of deformations, namely the volume strain which changes the volume of the
  system isotropically, and the shear strain which preserves the volume but changes the
  shape of the container.}
Under {\color{black} volume strains}, glasses melt by decompression,
and in presence of a hard-core repulsion, as in granular matter and in colloids,
they exhibit jamming upon compression.
The melting and the jamming transitions delimit the line where the glass remains solid.
Taking a glass on that line, one can probe its stability {\color{black} along the other axis of deformation, i.~e.} shear strain. 
Typically, the response of a glass to shear can be either (i)~purely elastic and stable (note that this does not mean that the response
is purely affine, as elasticity can emerge even in presence of a non-affine response), (ii)~partially plastic~\cite{schuh2007mechanical,rodney2011modeling,barrat2011heterogeneities}, which is {\color{black} accompanied by slip avalanches} and might be 
associated to the property of marginal stability~\cite{LN10,muller2015marginal}, or (iii)~purely plastic and unstable, once yielding takes place~\cite{rodney2011modeling,bonn2017yield,AFMB17}. 
{\color{black} Furthermore, granular materials~\cite{bi2011jamming} and dense suspensions~\cite{peters2016direct} may (iv)~jam when they are sheared.}

A question related to stability is \emph{reversibility}, i.e. to what extent a glass can recover its initial configuration when the deformation is released.
This question has been one of the key interests in cyclic shear experiments of colloidal suspensions~\cite{hyun2011review}. {\color{black} In simulations of some model glasses,}  it has been found that a reversible-irreversible transition accompanies the occurrence of yielding~\cite{regev2015reversibility,kawasaki2016macroscopic,leishangthem2017yielding}. 

The purpose of this work is to study, through extensive numerical  simulations, the {\color{black} volume}- and shear-strains phase diagram
of a model glass former, hard spheres (HS), {\color{black} in order
  to unify the above mentioned phenomena, i.e., plasticity, yielding, compression- and shear-jamming, and the failure of reversibility.
Thanks to the swap algorithm, introduced by Kranendonk and Frenkel~\cite{kranendonk1991computer} and recently adapted to simulate polydisperse HS systems with unprecedented efficiency~\cite{BCNO2016PRL}, we are able to prepare initial {\color{black}\it{equilibrium}} supercooled liquid configurations {\color{black} up to high densities going even beyond experimental limits~\cite{berthier2017breaking}.} {\color{black} While the standard molecular dynamics (MD) algorithm mimics the real dynamics, the swap algorithm accelerates the relaxation by introducing artificial exchanges of particles
  at different positions. With this trick a dense supercooled liquid state
  with very large relaxation time can be prepared.}
Given such a system,  by turning off
the swap moves and switching to standard MD simulations, the system is effectively confined in a glass state, because its relaxation time is much larger than the achievable MD simulation time. Perturbing this initial equilibrium state with a given rate of
compression, decompression or shear strain during MD simulations,
the system is driven out of {\color{black} the original equilibrium supercooled liquid state.}  In this way we study the out-of-equilibrium response  to these external perturbations of the glass  selected by the initial supercooled liquid configuration, 
thus realizing what in Ref.~\cite{RUYZ2015PRL} is called adiabatic {\it state following}.
Using this procedure,
we completely map out the degree of stability of the HS glasses corresponding to widely different preparation protocols.
We show that there is a unique mapping between different types of stability and reversibility,
that the stable and the marginally stable glass phases can be well separated by 
sensitive measurement protocols~\cite{BCJPSZ2016PNAS, jin2017exploring},
and that marginality is manifested by a new type of reversibility which we denote as {\it partial irreversibility}. 
}

The idea of establishing a  phase diagram to unify the glass transition,
jamming and yielding of amorphous solids was initially proposed by \FZ{Liu and Nagel~\cite{liu1998nonlinear, LN10},
and subsequently explored by others, see e.g. Refs.~\cite{trappe2001jamming, ikeda2012unified}.}
{\color{black} Here we explicitly construct such a phase diagram for HS glasses,  represented by a {\it stability-reversibility map}, 
which complements the conjecture in~\cite{LN10} with new ingredients, 
namely the existence of the marginal glass phase and the dependence on the quality of annealing~\cite{RUYZ2015PRL,biroli2017liu,UZ17}. 
Our phase diagram} is expected to  be reproducible in experiments on vibrated granular glasses~\cite{candelier2009creep,seguin2016experimental}
and on colloids~\cite{bonn2017yield}, {\color{black} while molecular glasses are usually described by soft potentials,
for which the phase diagram needs to be modified.}

The plasticity of amorphous solids has been extensively studied both in phenomenological~\cite{FL98,HKLP11,AFMB17, LW15} and first-principle~\cite{CKPUZ17,UZ17} theories.
{\color{black} According to the exact mean-field (MF) solution of the HS model in infinite dimensions~\cite{UZ17},
  { \color{black} the glass phase can be decomposed into stable regions where plasticity is absent, 
    and marginally stable regions where it is expected. The two phases are separated}
  by \YJJJ{a  line} where the so-called Gardner transition  takes place~\cite{CKPUZ17, BCJPSZ2016PNAS, jin2017exploring}}. {\color{black}
Determining whether this MF Gardner transition is also present in three dimensions is an
extremely hard and currently open problem.
Numerical simulations in three dimensions have found consistent evidence that a HS glass changes from a stable state to a marginally stable state across a certain threshold density before reaching jamming~\cite{BCJPSZ2016PNAS,jin2017exploring}, 
but are not capable to determine whether such a change corresponds to a phase transition or a crossover, due to the lack of {\color{black} a careful analysis of finite-size effects}.
Here we relate  the signatures of the Gardner transition/crossover
to the emergence of plastic behavior and avalanches~\cite{franz2016mean,muller2015marginal, BU2016NP}, which can be measured in simulations via the onset of partial plasticity and the emergence of 
a protocol-dependent shear modulus \cite{jin2017exploring, YZ2014PRE}. The Gardner threshold determined in this approach is consistent with an independent estimate based on the growth of a {\color{black} spin-glass--like} susceptibility~\cite{BCJPSZ2016PNAS}.
Because the scope of our work is not to decide on the existence of a sharp Gardner phase transition, here we keep the conventional use of the terminology ``Gardner transition", but do not exclude the possibility that it may become a crossover in three dimensions. Moreover, it remains an open question if the Gardner transition and the associated marginality is  of relevance to \YJJJ{other systems.} For example, 
\FZ{the absence of marginality has been reported in simulations of a three-dimensional soft-potential model~\cite{PhysRevLett.119.205501},
and a system of hard spheres confined in a one-dimensional channel~\cite{hicks2017gardner}.}
While details may change {\color{black} among various systems}, the approach used in this study provides an example of how to construct a stability-reversibility map for {\color{black} generic glasses}.
}

\begin{figure*}[t]
      \centerline{\includegraphics[width=\columnwidth]{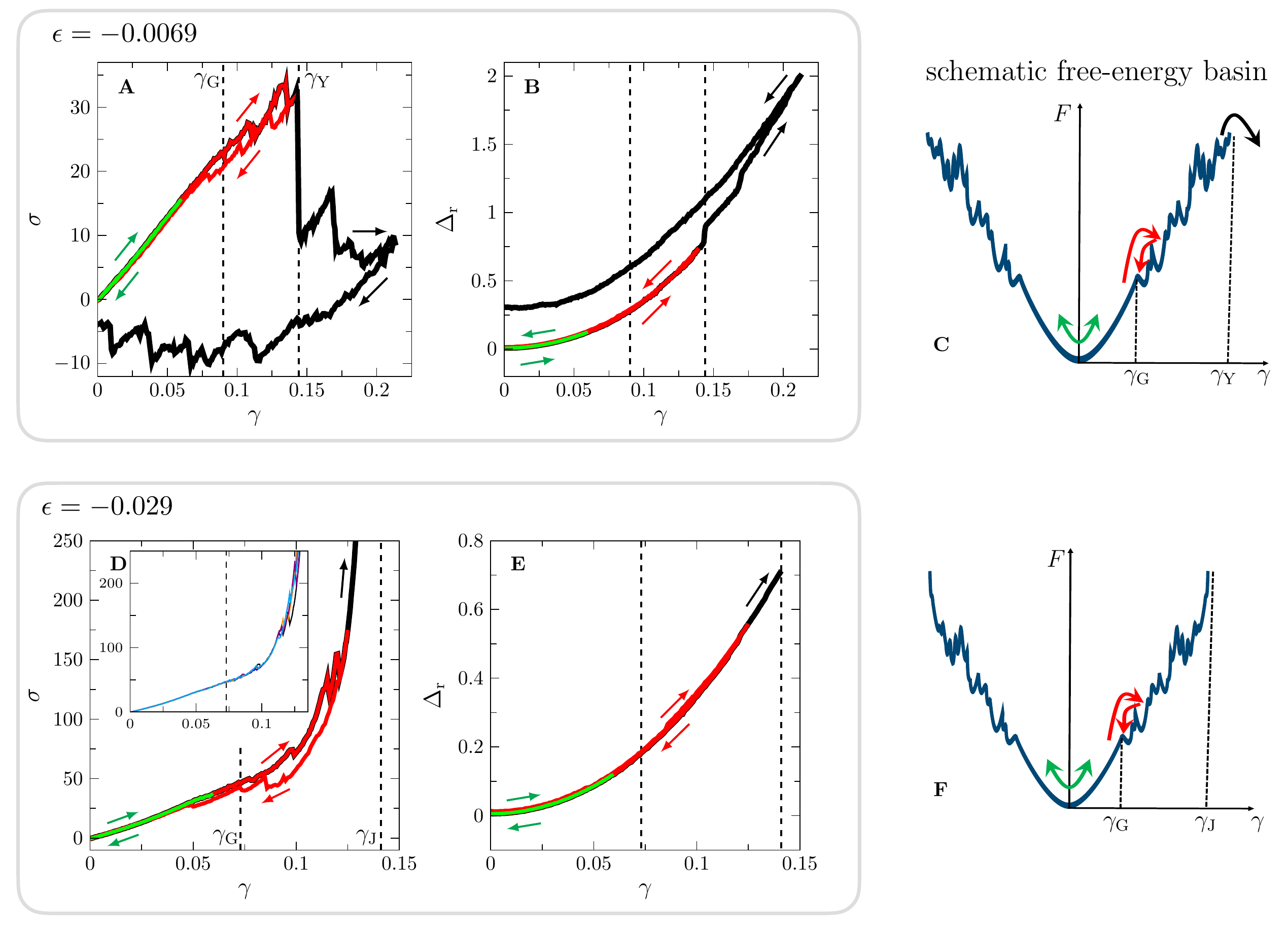} }
\caption{{\bf Reversibility, partial irreversiblity and irreversiblity of the HS glass under simple shear.}
Here we display typical behaviors of a glass sample 
obtained by annealing
up to $\varphi_{\rm g}= 0.655$.
{\bf (A)} Single-realization stress-strain curve of a glass at the fixed {\color{black} volume strain} \YJ{$\epsilon=-0.0069$} ($\varphi = 0.66$).
The shear strain is reversed at $\gamma = 0.06$ (green), 0.14 (red), and 0.2 (black). The smooth and jerky regimes are separated by $\gamma_{\rm G} \sim 0.09$. The yielding strain $\gamma_{\rm Y} \sim 0.144$ is also indicated. {\bf (B)} Corresponding plot of the relative mean squaredd displacement  $\Delta_{\rm r }$ as a function of $\gamma$.  {\bf (C)} {\color{black} Schematic} illustration of the free-energy glass basin under shear.
{\bf (D-F)} The same as {\bf (A-C)} but at  {\color{black} volume strain} \YJ{$\epsilon=-0.029$} ($\varphi = 0.675$)
for which $\gamma_{\rm G} \sim  0.073$ and $\gamma_{\rm J} \sim 0.14$.
In {\bf D}-inset, five different realizations for the same sample are plotted, \YJ{showing that plastic avalanches only occurs above $\gamma_{\rm G}$.}
See {\color{black} Fig.~S1} for the case $\epsilon=0.057$ ($\varphi = 0.62$) for which the system
      does not go through the partially irreversible regime under shear
      up to the yielding. {\color{black} The three cases ($\epsilon=-0.0069, -0.029, 0.057$) are indicated by black arrows in the stability-reversibility map~Fig.~\ref{fig:PD}.}
}
\label{fig:reversibility}
\end{figure*}

\section*{RESULTS}
\subsection*{Preparation of annealed glasses}

We study a three dimensional HS glass with continuous
polydispersity, identical to the one in Ref.~\cite{BCNO2016PRL} (see Materials and Methods).
Note that for HS, the temperature is irrelevant: it only fixes the overall kinetic energy of the system, which is related to the
sphere velocities, and thus to the unit of time. In our simulations, we set $k_{\rm B}T$ to unity. 
The relevant control parameters in this study 
are the packing fraction $\f$ and the shear strain $\g$. The reduced or dimensionless
pressure $p = P/(k_{\rm B}T \rho)$, with $P$ being the pressure and $\rho$ the number density, can be determined uniquely from the EOS  for given $\f$ and $\g$. 
Because the jamming limit is the point where the reduced pressure \FZ{of hard spheres} diverges, 
it corresponds, \FZ{for our system}, to the infinite pressure limit for fixed temperature, or the zero temperature limit for fixed pressure. 


One can consider HS as the limit of soft repulsive particles when 
 the interaction energy scale divided by $k_{\rm B}T$ goes to infinity: then, the HS system formally corresponds to the zero temperature
limit of soft repulsive particles in the unjammed phase where particles do not overlap. The jamming limit coincides in both systems, but the over-jammed phase is inaccessible by definition for HS. As a consequence, one of the axis (the temperature axis) in the Liu-Nagel phase diagram~\cite{LN10}
will be missing in our context. In fact, the HS phase diagram established here should correspond to the zero-temperature plane of the Liu-Nagel phase diagram without the over-jammed part. 

Our HS model is chosen in such a way that the particle swap moves~\cite{kranendonk1991computer}
can be used in combination with standard event-driven MD
to fully equilibrate the system up to very high densities,
covering a very
wide range of time scales for the standard MD dynamics without swap~\cite{BCNO2016PRL}.
Switching off the swap movements at volume fraction 
$\varphi_{\rm g}$ and leaving only MD acting on the particles
one gets effectively a HS amorphous solid, corresponding to the glass that would be formed during an annealing process that falls out
of equilibrium at $\varphi_{\rm g}$.
Therefore $\varphi_{\rm g}$ is the {\it glass transition density}. 
Because the system is still in equilibrium at $\varphi_{\rm g}$, its reduced pressure $p_{\rm g}$ 
follows the liquid equation of state (L-EOS) $p_{\rm g}=p_{\rm liq}(\varphi_{\rm g})$.

The possibility to explore a wide range of glass transition densities, thanks to the swap algorithm, is crucial to our work.
 In the following  we choose to work on three different values of  $\varphi_{\rm g}$, representing  ascending levels of annealing: 
\begin{itemize}
\item[(1)] {\bf Weakly annealed case:} $\varphi_{\rm g} = 0.609$, corresponding to the pressure $p_{\rm g} = 25.9$. 
Ref.~\cite{berthier2017breaking} fitted the data of $\alpha$-relaxation time $\tau_\alpha$ as a function of $p$ in liquids using \FZ{the standard Vogel-Fulcher-Tammann (VFT) form 
$\tau_\alpha = \tau_{\infty} \exp [A/(p_{\rm vft} -p)]$, a generalised VFT form $\tau_\alpha = \tau_{\infty} \exp [A/(p_{\rm vft} -p)^2]$,
and the facilitation model (FM) form  $\tau_\alpha = \tau_{\infty} \exp [A(p-p_{\rm fm})^2]$
(see~\cite{berthier2017breaking} for details on the fitting). 
We} estimate that the $\alpha$-relaxation time corresponding to $p_{\rm g} = 25.9$ is about $\tau_{\alpha}/\tau_0 \sim 5 \times 10^4$ \YJJJ{for all these forms},
where $\tau_0 \sim 10^4$ is the $\alpha$-relaxation time at  the onset density $\varphi_0 \approx 0.56$ of glassy dynamics.
Both VFT and FM forms give consistent values of $\tau_\alpha$. 
The time scale $\tau_\alpha/\tau_0 \sim 5 \times 10^4$ corresponds to a typical time scale \YJ{measured experimentally  in} colloidal glasses ($\tau_\alpha/\tau_0 \lesssim 10^5$ and $\tau_0 \approx 10^{-1}$s).

\item[(2)]{\bf Moderately annealed case:}  $\varphi_{\rm g} = 0.631$ and $p_{\rm g} = 30.9$. At this density,
\FZ{the standard VFT fitting gives
an estimated time scale $\tau_\alpha/\tau_0 \sim 3 \times 10^{10}$, the generalised VFT gives} $\tau_\alpha/\tau_0 \sim 10^{10}$, and the FM fitting gives $\tau_\alpha/\tau_0 \sim 10^{9}$. Such time scales are typically reachable in molecular glass forming liquids ($\tau_\alpha/\tau_0 \lesssim 10^{13}$ and $\tau_0 \approx 10^{-10}$s).

\item[(3)]{\bf Deeply annealed case:}  $\varphi_{\rm g}=0.655$ and $p_{\rm g} = 40.0$. At this density, the relaxation time is enormously large, and both VFT and FM fittings are unreliable. Ref.~\cite{fullerton2017density} measured the stability ratio $\mathcal{S}$ (the ratio between the melting time and the equilibrium relaxation time at the melting temperature) of this system. According to the data in~\cite{fullerton2017density}, the stability ratio at this density is around $\mathcal{S} \sim 10^3 - 10^5$ (the value depends on the melting pressure), which is comparable to experimental scales $\mathcal{S} \sim 10^2 - 10^5$ of  vapor-deposited ultra-stable glasses~\cite{singh2013ultrastable}. 
\end{itemize}
{\color{black} While the time scales we can access correspond to different materials, as discussed above, it is important to stress that molecular glass forming
liquids and ultra-stable glasses do not display a hard-core repulsion. The repulsion between molecules in these systems is usually better described by a Lennard-Jones--like soft potential. Therefore, some of the phenomena that we will describe in the following, which are strongly related to the presence of a hard-core potential,
will be absent in these materials. The most important example is jamming, which is by definition not present in Lennard-Jones--like soft potentials. The nature of the Gardner
transition could be also markedly different in some soft materials~\cite{PhysRevLett.119.205501}, and the applicability of some of our results on partial irreversibility
should then be checked. Yet, we believe that the HS model is a remarkable benchmark as it displays many important instability mechanisms (melting, yielding, compression- and shear-jamming, and the onset of marginal stability). It thus allows us to study in full details the interplay between these instability mechanisms and their dependence
on the quality of  annealing.
}

\subsection*{Stability and reversibility}

{\color{black}
Starting from the equilibrated supercooled liquid configurations at $\varphi_{\rm g}$, we now turn off the swap moves. Doing this, the liquid
relaxation time  goes beyond the time scale that we can access in our numerical experiments, and the system is thus effectively trapped into a glass
state.
We can then
follow the quasi-static evolution of the system under slow
changes of the {\color{black} volume strain}  $\epsilon = (\varphi_{\rm g} - \varphi)/\varphi$ and the shear strain $\gamma$ (see Materials and Methods),
and measure the corresponding evolution of the pressure and the shear stress. Although the system is formally out-of-equilibrium (from the liquid
point of view), one can reach a perfectly stationary state on the time scale we explore, restricted to the glass basin~\cite{RUYZ2015PRL}. 
The basin can then be followed
in restricted metastable ``equilibrium". We call the resulting trajectory in control parameter space metastable EOS or glass equations of state (G-EOS),   to distinguish it from the liquid equation of state (L-EOS). The G-EOS can be obtained
 by plotting the pressure and stress as functions of the {\color{black} volume} and shear strains.}

{\color{black} The change of {\color{black} volume strain} $\epsilon$ can be converted to that of volume fraction $\varphi$, via the relation 
$\varphi=\varphi_{\rm g}/(1+\epsilon)$.}
To achieve a change in {\color{black} volume fraction}, all particle diameters $D$ are uniformly changed
with rate $\dot{D}/D= 2 \times 10^{-4}$
for compression
and $\dot{D}/D= -2 \times 10^{-4}$
for decompression. 
The resulting rate of change of {\color{black} volume strain} is
$\dot{\epsilon} = -3(1+\epsilon)\frac{\dot{D}}{D}$.
The change of the shear strain is given at a rate $\dot \gamma  = 10^{-4}$.
The corresponding time scales of these rates
are in between the fast $\beta$- and the $\alpha$-relaxation times,
in such a way that the glass is followed nearly adiabatically, while the $\a$-relaxation remains effectively frozen~\cite{BCJPSZ2016PNAS,jin2017exploring}.
The target strains $(\epsilon,\gamma)$ can be achieved starting from the initial point $(0,0)$ following various paths in the {\color{black} volume}-shear strain plane. For example one can apply first a shear strain followed by a {\color{black} volume strain} or vice versa.
In the following we specify explicitly the paths that we follow and check the dependency of the final outcome on the choices of paths.

{\color{black}Now let us start our analysis by considering} what happens following a simple cyclic deformation:
first, the system is strained normally
$(0,0) \to (\epsilon,0)$, then sheared $(\epsilon,0) \to (\epsilon,\gamma)$
and finally sheared back in the reversed way $(\epsilon,\gamma) \to (\epsilon,0)$.
The following three typical behaviors are found: the response of the glass can be reversible, partially irreversible,
or totally irreversible, which {\color{black} signals} stable, marginally stable,
and unstable states {\color{black} of the glass}.
Typical examples of the stress-strain curves are shown in Fig.~\ref{fig:reversibility}.

(i) {\bf Reversible regime}: For small $\gamma$, the stress $\sigma$ increases smoothly and monotonically with increasing $\gamma$
(green lines in Fig.~\ref{fig:reversibility}A and D). To the first order the stress is linear,  $\delta \sigma = \mu \delta \gamma$, where $\mu$ is the shear modulus. If the strain is released with $-\dot \gamma$,
the stress-strain curve reverses to the origin --  this is a typical elastic response.

(ii) {\bf Partially irreversible regime}:
For larger $\gamma$ \YJ{above a certain threshold $\gamma_{\rm G}$}, the stress-strain curve becomes jerky, consisting of piecewise linear elastic responses followed by small but abrupt stress drops {\color{black} (red lines in Fig.~\ref{fig:reversibility}A and D)}. Each stress drop corresponds to a plastic event, where some particles rearrange their positions. The glass in this regime is marginally stable in the sense that a tiny $\delta \gamma$ could make the system unstable by triggering such plastic events, but the particles immediately find another stable configuration \textcolor{black}{nearby} avoiding further failure of the entire system. \YJ{Although the stress-strain curve is locally irreversible for small reversed strain, globally it eventually {\color{black} returns} to the origin  when the shear strain is released back to $\g=0$ {\color{black}(the red lines in Fig.~\ref{fig:reversibility}A and D merge with the green lines below $\gamma_{\rm G}$)}. We call such behavior partial irreversibility.}

(iii) {\bf Limit of existence of the solid}:
For even larger $\gamma$, the glass faces two kinds of consequences depending on the {\color{black} volume strain} $\epsilon$ applied before shearing.
\begin{itemize}
\item {\bf Yielding}: \YJ{At the yielding strain $\gamma_{\rm Y}$,} a sudden and significant stress drop occurs. When this happens, the entire system breaks into two pieces that can slide with respect to each other along a fracture. As shown by the stress-strain curve (black line in Fig.~\ref{fig:reversibility}A), yielding is irreversible -- once the glass is broken, it can not be ``repaired".
In a costant volume protocol where we keep the total volume of the system unchanged, yielding can be seen only if the system is not compressed to too
high packing fractions, i.e. for not too negative {\color{black} volume strain} $\epsilon$.
\item {\bf Shear jamming}: The behavior changes dramatically if the system is compressed more before shearing. In this case  the system jams at \YJ{the shear jamming strain $\gamma_{\rm J}$}, which is signalled by the divergence of the shear stress (black line in Fig.~\ref{fig:reversibility}D).
  \end{itemize}

To examine the reversibility more carefully, we measure the relative mean squared displacement  $\Delta_{\rm r }$ (see Materials and Methods for the definition) between the initial state at $\gamma=0$ before the shear is applied, and the final state at $\gamma=0$ after a single cycle of shear is applied (Fig.~\ref{fig:reversibility}B and E). 
If the initial and the final configurations are identical, $\Delta_{\rm r} =0$; otherwise, the more different they are, the larger $\Delta_{\rm r}$ is.
The value of $\Delta_{\rm r }$ returns to zero in the reversible and partially irreversible cases, but becomes non-zero in the irreversible case, being consistent with the above analysis based on the stress-strain curves. Note that here we neglect differences on  the microscopic scale of vibrational cage size $\Delta\lesssim 0.01$ (see Materials and Methods), i.e., \YJ{a system is {\color{black}called} irreversible only if the difference on $\Delta_{\rm r }$ between the initial and final configurations is larger than $\Delta$. We have also examined that the above behaviors persist in multi-cycle shears (see {\color{black} Fig.~S2}). }

It is useful to understand our observations using a schematic picture of the free-energy landscape. Each glass state  is represented by a basin of free-energy $F(\varphi_{\rm g}; \epsilon, \gamma; N)$, which is distorted upon increasing shear strain $\gamma$ (Fig.~\ref{fig:reversibility}C and F).
The shear stress is nothing but the slope of the free-energy \YJ{$\sigma(\varphi_{\rm g}; \epsilon, \gamma; N)= \frac{\beta}{N}\frac{\partial F(\varphi_{\rm g}; \epsilon, \gamma; N)}{\partial \gamma}$} with $\beta$ being the inverse temperature.
{\color{black} The associated shear modulus is obtained
by taking one more derivative with respect to $\gamma$,
which gives nothing but the curvature of the free-energy basin.}
In the stable regime, the basin is smooth;
in the marginally stable regime, the basin becomes rough, consisting of many sub-basins {\color{black} with larger associated shear modulus},
which results in the failure of pure elasticity~\cite{HKLP11,YZ2014PRE,BU2016NP}. In this state, the system can release the stress 
via hopping between different sub-basins, \YJ{corresponding to plastic events},
  {\color{black} which leads  to emergent slow relaxation of shear stress}
\cite{YZ2014PRE,jin2017exploring}.
For very large strains the system either yields by escaping from the glass basin (Fig.~\ref{fig:reversibility}C) or jams by hitting the vertical wall due to the hard-core constraint (Fig.~\ref{fig:reversibility}F).

The plastic behavior appearing in the partially irreversible regime is taking place at mesoscopic scales, and it would be averaged out in a macroscopic system at large enough time scales~\cite{YZ2014PRE}. 
There is evidence which shows  that the minimum strain 
increment $\delta \gamma_{\rm trigger}(N)$ to trigger a plastic event
vanishes in the thermodynamic limit $N \to \infty$~\cite{HKLP11,karmakar2010statisticalB}. 
This implies that in a macroscopic system, any small but finite increment of strain would cause a non-zero number of mesoscopic plastic events~\cite{muller2015marginal}.
\YJ{Moreover, time-dependent aging effects associated to such plastic events were observed in stress relaxations~\cite{jin2017exploring}.}
Therefore, in macroscopic systems at large enough time scales the plasticity would be averaged out,  and one would observe just a renormalized  ``elastic" response. The bare elastic response can only be seen within the piece-wise linear mesoscopic response for $\delta \gamma < \delta \gamma_{\rm trigger}(N)$.
This means that two different shear moduli can be defined: the bare one
$\mu_{\rm bare}=\lim_{N \to \infty} \lim_{\delta \gamma \to 0} \delta \sigma(\varphi_{\rm g}; \epsilon, \gamma; N)/\delta \gamma$ that takes into account the piecewise elastic behavior between two subsequent avalanches,  and the macroscopic one
$\mu_{\rm macro}=\lim_{\delta \gamma \to 0} \lim_{N \to \infty}\delta \sigma(\varphi_{\rm g}; \epsilon, \gamma; N)/\delta \gamma$, which represent the average behavior and is smaller than the former~\cite{HKLP11,YZ2014PRE}. 
Therefore the small strain $\d\g\to 0$ limit and the thermodynamics limit do not commute in the marginal plastic phase 
(see Text S1 for a detailed discussion).

\begin{figure*}[t]
  \centerline{\includegraphics[width=0.55\columnwidth]{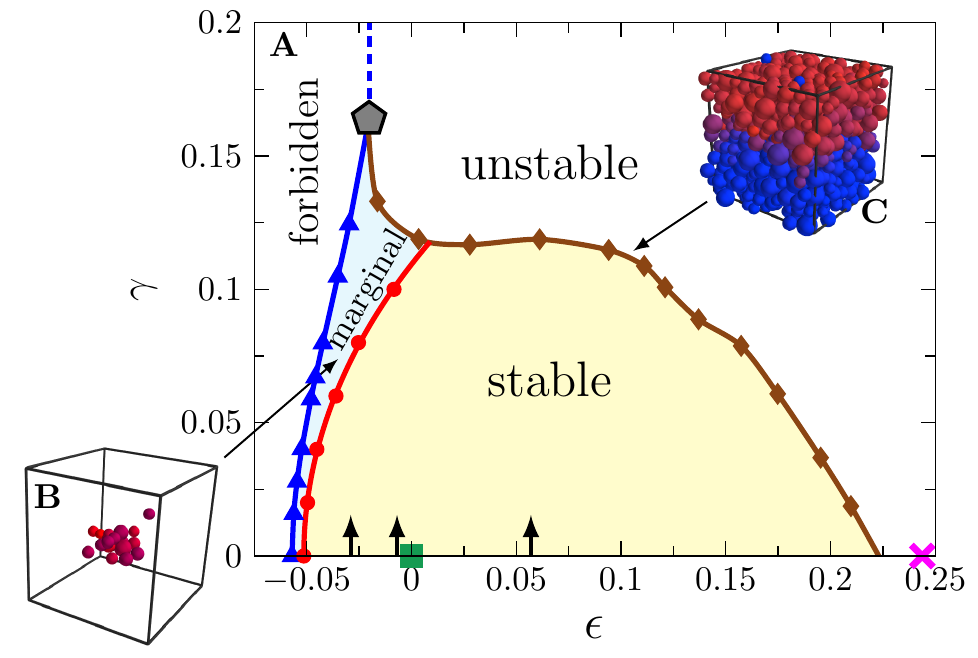} }
\caption{
  {\bf Stability-reversibility map.}
{\bf (A)}  Stability-reversibility map of the HS glass annealed up to  $\varphi_{\rm g} = 0.655$, 
obtained by  the constant pressure-shear (CP-S) protocol {\color{black}(see Materials and Methods for the definition)}.
The horizontal axis is the {\color{black} volume strain} $\epsilon$ and
  the vertical axis is the shear strain $\gamma$. The origin $(0,0)$ marked by the
  green square represents the initial glass without deformations.
   The glass remains stable only inside a region
   bounded by the yielding line $\gamma_{\rm Y}(\varphi_{\rm g};\epsilon)$ (brown diamonds) and the shear-jamming line $\gamma_{\rm J}(\varphi_{\rm g};\epsilon)$ (blue triangles). The shear-yielding and the shear-jamming lines meet at the yielding-jamming crossover point (gray pentagon) \YJ{($\epsilon_{\rm c}, \gamma_{\rm c})=(-0.020(4), 0.16(1))$} (corresponding to $\varphi_{\rm c} = 0.669(3)$). The Gardner line $\gamma_{\rm G}(\varphi_{\rm g}; \epsilon, \gamma)$ (red circles) separates the marginally stable glass phase (blue area) from the stable glass phase (yellow area). Under decompression, i.e., increasing $\epsilon$ with $\gamma=0$ the glass becomes fully liquified at the melting point (pink cross). In the plot, the stable, marginally stable and unstable regimes correspond to reversible, partially irreversible and irreversible regimes respectively. 
   {\color{black} The black arrows at the bottom indicate the volume strains
     used in Fig.~\ref{fig:reversibility} (the two arrows on the left of the green square 
at  $\epsilon=-0.0069$ and $\epsilon=-0.029$, {\color{black} which are above and below $\epsilon_{\rm c}$ respectively})
      and Fig.~S1 (the arrow on the right at $\epsilon = 0.057$).}
   {\bf (B)} Snapshot shows the particles involved in a typical plastic event
   in the marginally stable glass phase.  {\bf (C) } Snapshot shows the planar fracture structure that appears during yielding. The colors in {\bf (B-C)} represent the relative single particle displacement $\delta_{\rm r}^i = \sqrt{\Delta_{\rm r}^i}$; warmer colors indicate higher values.
}
\label{fig:PD}
\end{figure*}

\begin{figure*}[h]
\centerline{\includegraphics[width=0.9\columnwidth]{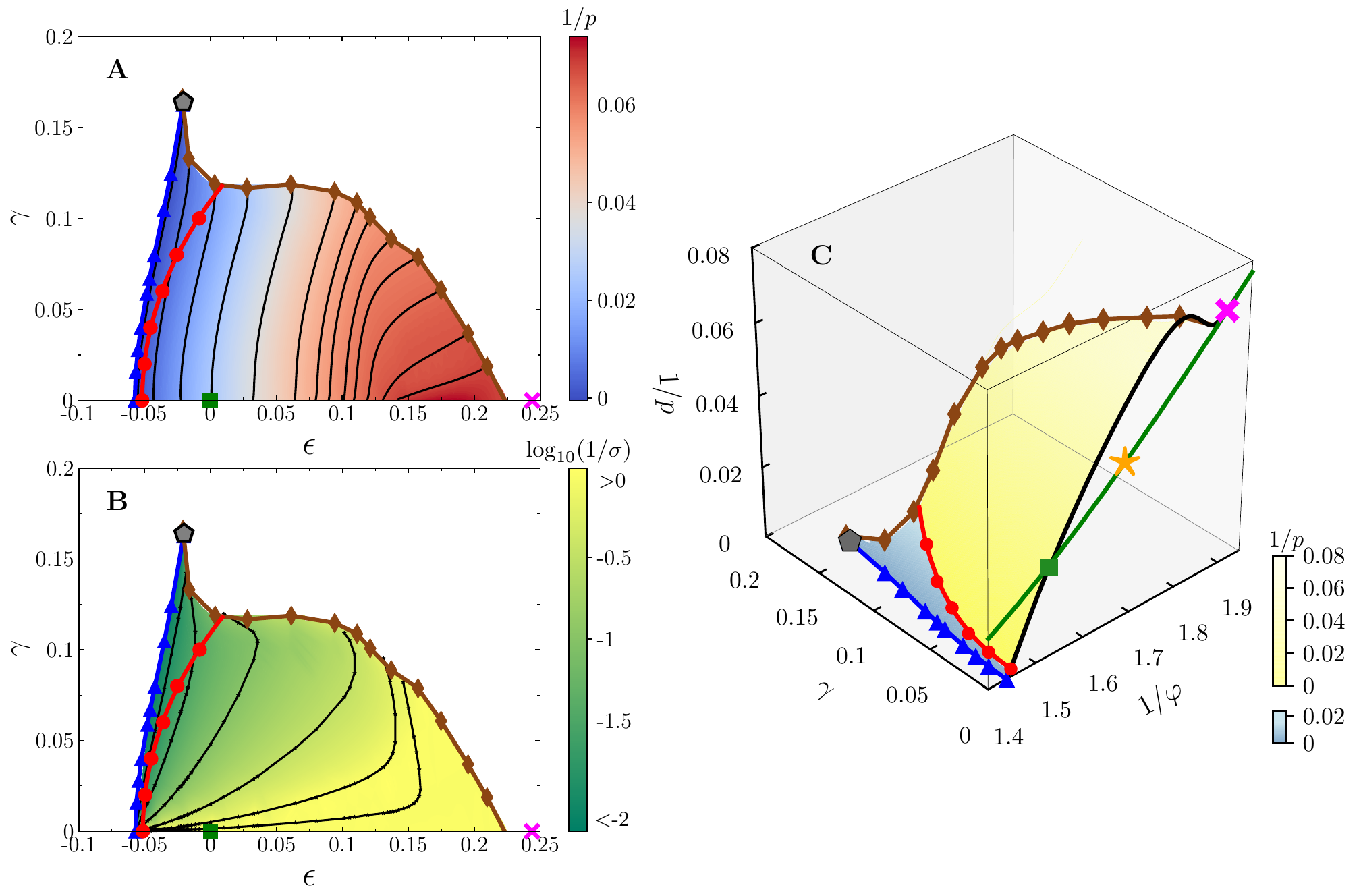} }
\caption{
 {\bf  Glass equations of state.}
{ \bf (A) and (B)} We show the G-EOSs for the pressure $p=p_{\rm glass}(\varphi_{\rm g};\epsilon,\gamma)$ and shear stress $\sigma=\sigma_{\rm glass}(\varphi_{\rm g};\epsilon,\gamma)$ for the HS glass prepared at density
 $\varphi_{\rm g} = 0.655$ by heat maps.
The color bar scales represent $1/p$ and {\color{black} $\log_{10}(1/\sigma)$}. \YJ{The thin black lines represent isobaric (constant-$p$) lines for $p=$ 14.5, 15.0, 15.8, 16.5, 17, 18, 19, 21, 27, 40, 65, 160, 1000 (from right to left) in {\bf(A)}, and constant-$\sigma$ lines for $\sigma = 0.3, 1, 3, 10, 30, 100$ (from right to left) in {\bf(B)}.}
The data are obtained via the CP-S protocol
 {\color{black}(see Materials and Methods for the definition)}.
{\bf (C)} Three dimensional  view of the same G-EOS for the pressure
(colored plane)  and the L-EOS $p=p_{\rm liq}(\varphi)$ (green line). 
Here specific volume $1/\varphi$ 
is used instead of
the {\color{black} volume strain}  $\epsilon$.
The evolution of liquid under compression/decompression
follows the Carnahan-Stirling empirical liquid EOS~\cite{BCJPSZ2016PNAS}.
The golden star represents the mode-coupling theory (MCT) transition  point, which is obtained from extrapolation of the relaxation time according to the MCT  scaling~\cite{BCJPSZ2016PNAS}.
Note that $p_{\rm glass}(\varphi_{\rm g};0,0)=p_{\rm liq}(\varphi_{\rm g})$ holds by the definition of $\varphi_{\rm g}$. See Fig.~\ref{fig:PD} for the meaning of  symbols. See {\color{black} Fig.~S9} for the cases of other  $\varphi_{\rm g}$ and other exploration protocols.
}
\label{fig:G-EOS}
\end{figure*}

\subsection*{Stability-reversibility map and glass equations of state}

These three different kinds of responses of the system to simple cyclic shear,
{\color{black}listed above as (i)-(iii)},
can be summarized by the stability-reversibility map in the $\epsilon-\gamma$ plane as shown in Fig.~\ref{fig:PD}.
{\color{black} There we also show a typical plastic event in the marginal phase (Fig.~\ref{fig:PD}B) and a yielding event (Fig.~\ref{fig:PD}C), which clearly indicate two different mechanisms that can cause a failure of stability.}
As long as the glass remains stable \YJ{or marginally stable}, its macroscopic properties can be 
characterized by the {\color{black} G-EOS} for the pressure
$p=p_{\rm glass}(\varphi_{\rm g};\epsilon,\gamma)$ and
the shear stress $\sigma=\sigma_{\rm glass}(\varphi_{\rm g};\epsilon,\gamma)$
as shown in Fig.~\ref{fig:G-EOS}A and B.
\YJ{The pressure $p$ and the stress $\sigma$ are derivatives of the glass free-energy $-\beta F(\varphi_{\rm g}; \epsilon, \gamma)$ with respect to $\epsilon$ and $\gamma$ respectively. }


Along the $\gamma=0$ line, the evolution of the system under {\color{black} volume strain} $\epsilon$ will eventually lead the system 
either to jamming after sufficient compression $\epsilon < 0$ or melting after sufficient decompression $\epsilon > 0$.
At jamming, particles form an isostatic rigid contact network such that no further compression can be applied.
Decompressing the system reduces $p$, which eventually melts the system into {\color{black} a liquid state}.
The evolution of the pressure $p$ follows the zero-shear strain G-EOS $p=p_{\rm glass}(\varphi_{\rm g};\epsilon,0)$
both upon compression and decompression. Obviously $\sigma=\sigma_{\rm glass}(\varphi_{\rm g};\epsilon,0)=0$.

\YJ{Applying a shear strain at any point on the $\gamma=0$ G-EOSs $p=p_{\rm glass}(\varphi_{\rm g};\epsilon,0)$ and $\sigma=\sigma_{\rm glass}(\varphi_{\rm g};\epsilon,0)$ allows us to explore the {\color{black} volume}-strain versus shear-strain phase diagram and we can track both the pressure
$p=p_{\rm glass}(\varphi_{\rm g};\epsilon,\gamma)$ and the stress  $\sigma=\sigma_{\rm glass}(\varphi_{\rm g};\epsilon,\gamma)$.}
Under shear the glass has two possible fates: either it yields across the shear-yielding line $\gamma=\gamma_{\rm Y}(\varphi_{\rm g};\epsilon)$,
or it jams at the shear-jamming line $\gamma=\gamma_{\rm J}(\varphi_{\rm g};\epsilon)$.
Yielding can be detected by analyzing the stress-strain curve, i.~e. $\sigma_{\rm glass}(\varphi_{\rm g};\epsilon,\gamma)$ versus  $\gamma$, 
while shear jamming is signaled by a divergence of both the pressure
$p_{\rm glass}(\varphi_{\rm g};\epsilon,\gamma) \to \infty$ and the stress 
$\sigma_{\rm glass}(\varphi_{\rm g};\epsilon,\gamma) \to \infty$ as $\gamma \to \gamma_{\rm J}(\varphi_{\rm g};\epsilon)$.
The shear-yielding and the shear-jamming lines define the boundaries of the stability of the HS glass, beyond which the glass is unstable or simply forbidden.
\YJ{The two lines meet at  a {\it yielding-jamming crossover point} ($\epsilon_{\rm c}(\varphi_{\rm g}), \gamma_{\rm c}(\varphi_{\rm g})$), or ($\varphi_{\rm c}(\varphi_{\rm g}), \gamma_{\rm c}(\varphi_{\rm g})$).}

Within the boundary of the stability-reversibility map, there are two {\color{black} phases}:
the stable (reversible) phase, and
the marginally stable (partially irreversible) phase. We call the line which separates the two a {\it Gardner line}.
{\color{black} 
  Across this line the qualitative nature of the system's response to deformations changes: the stress-strain curve is smooth within the
  stable (reversible) phase 
  but jerky in the marginally stable (partially irreversible) phase.}
Interestingly, the stability-reversibility map shown in Fig.~\ref{fig:PD} suggests that if 
we choose an $\epsilon$ such that the Gardner line is not crossed along the path $(\epsilon, 0) \to (\epsilon, \gamma_{\rm Y})$, 
then no marginally stable region should be observed. 
{\color{black}Fig.~S1} shows such a case {\color{black}(with $\epsilon=0.057$)}
where we do not observe partial irreversibility all the way up to yielding. The term {\it Gardner line} is inferred from the MF glass theory~\cite{CKPUZ17,UZ17}, in which a continuous phase transition, 
\YJ{the Gardner transition},  occurs on this line.
   {\color{black}
 However, whether it is a genuine transition line or
     crossover line in three dimensions is an open question as we noted in the
     introduction. In the next sub-section we will explain how we estimate this
     line numerically in the present system.}

We made the choice in Fig.~\ref{fig:PD} to represent the stability-reversibility map in terms of strains ({\color{black} volume} and shear). {\color{black} In Fig.~S3A, we plot it in terms of volume fraction $\varphi$ and shear strain $\gamma$, which can be directly compared to the theoretical prediction in Ref.~\cite{UZ17}.
In} some experiments 
the shear stress is controlled instead of the shear strain, and in that case it is customary to represent the phase diagram
in the density-stress plane. Such a figure is reported in {\color{black} Fig.~S3B}, which is directly comparable to the phase diagram reported
in the granular experiment of Ref.~\cite{candelier2009creep}.

The stability-reversibility map and the G-EOS depend on the preparation density $\varphi_{\rm g}$ of the glass, 
which represents the depth of annealing. As shown in Fig.~\ref{fig:G-EOS}C, where the glass and liquid EOSs
are displayed together, \YJ{the $\gamma=0$ G-EOS and the L-EOS intersect at the point $(\varphi,\gamma)=(\varphi_{\rm g},0)$, which shows the intrinsic connection between glass and liquid EOSs.} The initial unperturbed glass is located at $(\epsilon,\gamma)=(0,0)$ in the stability-reversibility map.

\begin{figure*}[t]
  \centerline{\includegraphics[width=\columnwidth]{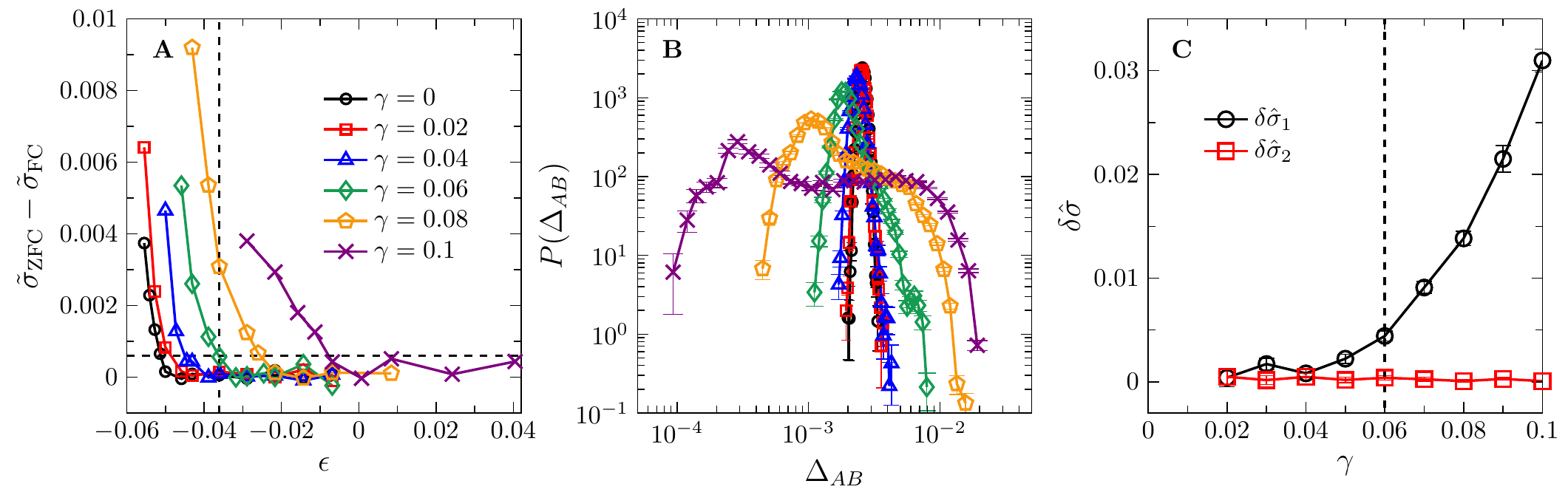} }
\caption{{\bf Marginal stability and  partial irreversiblity.} {\bf (A)} The difference between ZFC  and FC stresses (rescaled by $p$, $\tilde \sigma = \sigma/p$) as a function of the {\color{black} volume strain} $\epsilon$, for $\varphi_{\rm g} = 0.655$ and a few differently fixed $\gamma$. The Gardner threshold $\epsilon_{\rm G}(\varphi_{\rm g}; \gamma)$ is determined as the point where this difference exceeds 0.0006 (horizontal dashed line). For example, the vertical dashed line marks $\epsilon_{\rm G} (\varphi_{\rm g} = 0.655; \gamma = 0.06) \approx -0.036$. {\bf (B)} The distribution of $\Delta_{AB}$ over samples and realizations, at fixed {\color{black} volume strain}  $\epsilon=-0.036$ (or $\varphi = 0.68$), 
  for different $\gamma$. From the data, we estimate $\gamma_{\rm G} (\varphi_{\rm g} = 0.655; \epsilon = -0.036) \approx 0.06$, where $P(\Delta_{AB})$ becomes clearly non-Gaussian (green points), consisting with {\bf (A)}. {\bf (C)} The stress drops  $\delta \hat{\sigma}_1$  and  $\delta \hat{\sigma}_2$ measured in single cycle shear simulations are plotted as functions of $\gamma$ at fixed $\epsilon=-0.036$. {\color{black}(see the text for the definitions.)} The vertical dashed line represents $\gamma_{\rm G} (\varphi_{\rm g} = 0.655; \epsilon = -0.036) \approx 0.06$  estimated in {\bf (A) and (B)}. Note that at $\epsilon=-0.036$, the system jams under constant volume shear (see Fig.~\ref{fig:PD}) so that we can exclude any irreversibility caused by yielding.}
\label{fig:Gardner}
\end{figure*}

\subsection*{Marginal stability and partial irreversibility}

Having presented above our most important results,
in the following we show more details on how the stability-reversibility map and the G-EOS are obtained in our numerical experiments.  To this end, at each $\varphi_{\rm g}$, we prepare $N_{\rm s} = 100$ independent equilibrium supercooled liquid configurations by the swap algorithm, which have different equilibrium positions of particles, and are called {\it samples}.
By switching off the swap, they become glasses. For each sample of glass, we repeat $N_{\rm r} \sim 50 -200$ {\it realizations} of a given protocol which is a combination of  compression (or decompression) and simple shear. Each realization starts from statistically independent initial particle velocities drawn from the Maxwell-Boltzmann distribution at $\varphi_{\rm g} $.

The Gardner transition marks the point where the elastic behavior is replaced by a partially plastic one.
Avalanches and plasticity are extremely marked in finite size systems, while they are averaged out
on macroscopic length and time scales.
Furthermore in finite-size systems, even though each individual stress-strain curve is jerky in the marginal phase as shown in Fig.~\ref{fig:reversibility},  
 the average over different samples and realizations washes out all the sudden drops giving rise to a smooth profile.
Therefore macroscopic G-EOSs by themselves
do not allow the detection of the marginally stable phase (see Text S1 and Fig.~S4 for a detailed discussion). 
In order to precisely locate the onset of plasticity and the marginal phase, we will examine the
hysteretic response to very small shear increments.


Inspired from spin glass experiments~\cite{nagata1979low}, we compare the shear stress measured by two different protocols, the so-called zero field compression (ZFC) and the field compression (FC) protocols~\cite{jin2017exploring}.
Within the FC protocol, one first compresses the system and then shear it. In the ZFC one instead reverses the order (see Materials and Methods for more details). The FC stress  $\sigma_{\rm FC}$ can be considered as the large time limit of  $\sigma_{\rm ZFC}$, as long as the yielding and the $\alpha$-relaxation do not occur~\cite{YZ2014PRE, jin2017exploring}. For elastic solids such as crystals, the two stresses are identical. For marginally stable glasses, however,  $\sigma_{\rm FC}$ is lower than $\sigma_{\rm ZFC}$, because of the stress relaxation associated to the plastic events happening at mesoscopic scales.
 {\color{black} The origin of two responses can be attributed
  to the organization of free-energy landscape
  shown schematically in Fig.~\ref{fig:reversibility}C and F.
Roughly speaking, the ZFC stress $\sigma_{\rm ZFC}$
  is dominated by the short time response within the small sub-basins,
  while the FC stress $\sigma_{\rm FC}$ reflects the renormalized, long time response within the big envelope of sub-basins.}
The bifurcation point between the two stresses determines the Gardner point. {\color{black} Note that this criterion to determine the Gardner
 point is the same as the one used in Ref.~\cite{jin2017exploring}.}
 Figure~\ref{fig:Gardner}A shows the data used 
to obtain the Gardner points $\epsilon_{\rm G}(\varphi_{\rm g}; \gamma)$ for a few different values of $\gamma$. 
Connecting the Gardner points gives the Gardner line $\gamma=\gamma_{\rm G}(\varphi_{\rm g}; \epsilon)$
in Fig.~\ref{fig:PD}. See Fig. S5 for the same results obtained for other values of $\varphi_{\rm g}$.

Alternatively, one may look at caging order parameters such as the mean squared displacement $\Delta$ and
the typical separation between two replicas $\Delta_{AB}$~\cite{BCJPSZ2016PNAS} (see Materials and Methods for more precise definitions). 
The two replicas are generated from the same initial sample in two independent realizations. They are firstly compressed 
to a target $\epsilon$
under zero shear strain, and then sheared to the target shear strain $\gamma$ under the fixed $\epsilon$. When the Gardner point $\gamma_{\rm G}(\varphi_{\rm g}; \epsilon)$ is crossed over, $\Delta$ and $\Delta_{AB}$ should also separate. However this is a sign of critical behavior only if the corresponding susceptibility $\chi_{AB} = N \frac{\langle \Delta_{AB}^2\rangle - \langle \Delta_{AB}\rangle^2}{\langle \Delta_{AB}\rangle^2}$ grows~\cite{hicks2017gardner}. Here $\langle \ldots \rangle$ represents the average over both samples and realizations. $\chi_{AB}$ is a {\color{black} spin-glass--like} susceptibility whose growth suggests the increase of heterogeneity and cooperativity in the system as suggested by the MF theory~\cite{CKPUZ17}.
The behavior of $\chi_{AB}$ can be inferred from Fig.~\ref{fig:Gardner}B where we plot the probability distribution $P(\Delta_{AB})$ of $\Delta_{AB}$. We clearly see a Gaussian-like behavior below the Gardner threshold, fat-tailed around it, and double peaked above it. The Gardner point inferred in this way
$\gamma_{\rm G} (\varphi_{\rm g} = 0.655; \epsilon = -0.036) \approx 0.06$ 
is consistent with the determination from ZFC-FC protocols 
$\epsilon_{\rm G} (\varphi_{\rm g} = 0.655; \gamma = 0.06) \approx -0.036$.

This result provides a strong evidence that  partial irreversibility and plasticity in Fig.~\ref{fig:reversibility} are essentially related to emerging marginal stability.
We perform the following test to examine their connections more directly.
Starting from a compressed glass at $(\epsilon,\gamma=0)$, we first shear the glass to a target shear strain at $(\epsilon, \gamma - \delta \gamma)$ under constant volume, then apply an additional cycle of small shear strain $\delta \gamma = 0.004$, following the path $(\epsilon, \gamma  - \delta \gamma ) \to (\epsilon, \gamma  )  \to (\epsilon, \gamma  - \delta \gamma) $. If the system is reversible, then the difference between the stresses before and after the single cyclic shear, $\delta \hat{\sigma}_1 = (\sigma_{\rm before} - \sigma_{\rm after})/\sigma_{\rm before}$, should be zero, otherwise not. Figure~\ref{fig:Gardner}C confirms that $\delta \hat{\sigma}_1 (\gamma)$ begins to grow around the  $\gamma_{\rm G}$ estimated from the other two approaches 
described before (Fig.~\ref{fig:Gardner}A and B). 
However, such kind of irreversibility is only partial, because the system is reversible under a circle of shear with larger strain. Indeed, systems following the path $(\epsilon, \delta \gamma ) \to (\epsilon, \gamma )  \to (\epsilon, \delta \gamma ) $, where $\delta \gamma = 0.004$ is fixed and $\gamma$ is varying,  show that the stress difference $\delta \hat{\sigma}_2$ is nearly zero for any $\gamma$.

   {\color{black} 
Finally it is important to stress that in our three dimensional numerical simulations, as in previous ones~\cite{BCJPSZ2016PNAS,jin2017exploring}, 
we cannot decide on whether the separation between the stable and marginally stale phase corresponds to a true phase transition. This would
require, for instance, a careful study of finite size effects
on $\chi_{AB}$, to extract the behavior for
$N\to\io$, which is very difficult already in much simpler models such as
spin glasses. The focus of our work is on relating the Gardner line, which is only a (quite sharp) crossover in our simulations, to the onset
of partial irreversibility.
}

\begin{figure*}
   \centerline{\includegraphics[width=\columnwidth]{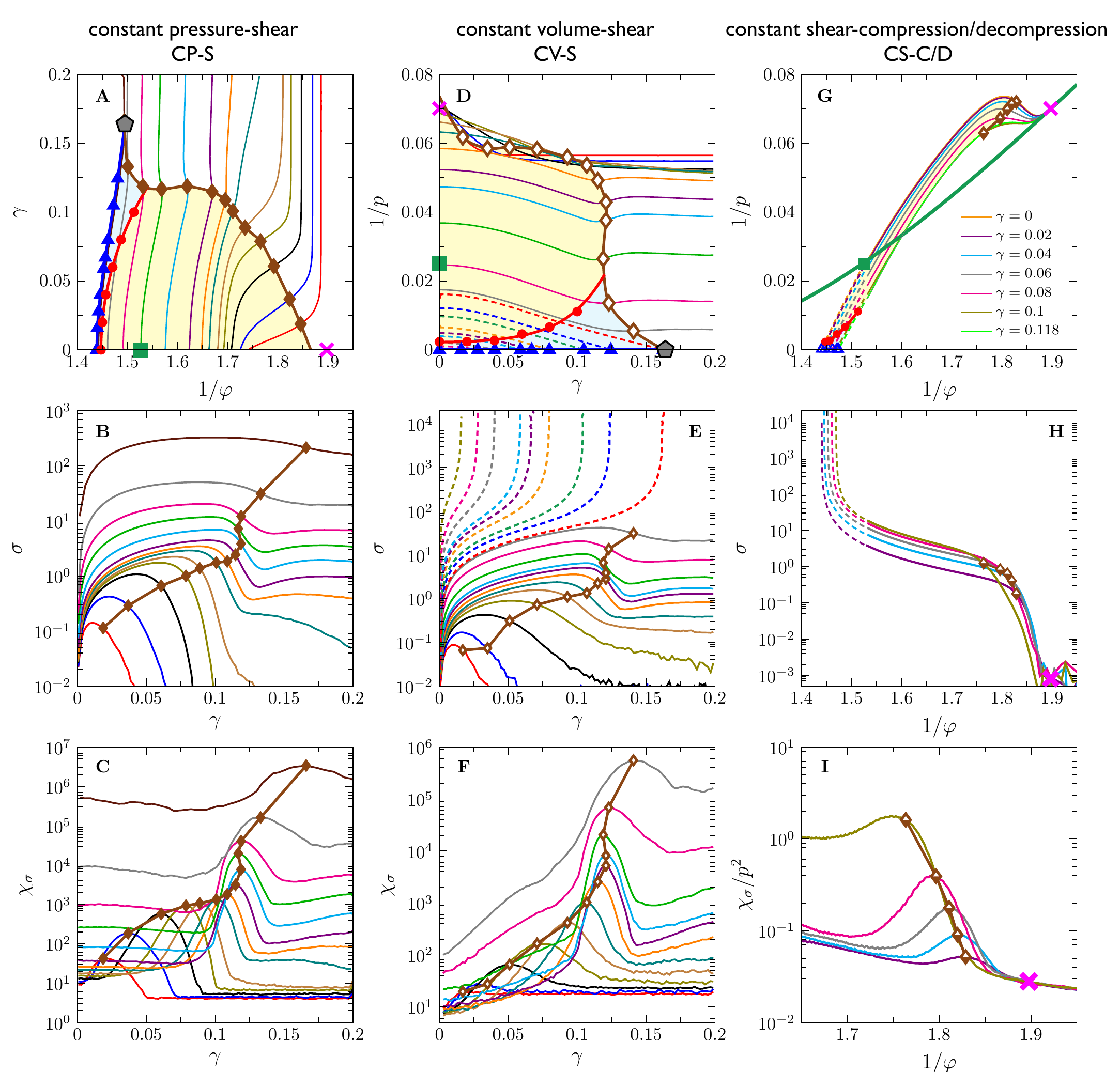} }
\caption{{\bf 
Yielding, shear jamming, and melting 
 in {\color{black} constant pressure-shear (CP-S), constant volume-shear (CV-S), and constant shear strain-compression/decompression ({\color{black}CS-C/D})}
 protocols.} 
 {\bf (A-C)}  G-EOSs obtained by the CP-S protocol for the $\varphi_{\rm g}=0.655$ system.
The solid thin  lines are isobaric lines for $p =  14.5, 15.0, 15.8, 16.5, 17, 18, 19, 21, 27, 40, 65, 160,  1000$ (from right to left in {\bf A}. The same colors are used for {\bf B} and {\bf C}.) 
The yielding line (filled brown diamonds) for the CP-S protocol are determined from the peak positions of  the stress susceptibility-strain ($\chi_\sigma - \gamma$) curves shown in {\bf (C)}. {\bf (D-F)} G-EOSs obtained by the CV-S protocol.
The solid thin lines are  isochoric (constant-$\varphi$) lines for  $\varphi = 0.558, 0.562, 0.568, 0.574, 0.579, 0.586, 0.595, 0.607, 0.616, 0.635, 0.655, 0.667$, where the system exhibits yielding (from bottom to top in {\bf E}).
The dashed thin lines are isochoric lines  for $\varphi = 0.669,  0.676, 0.679, 0.684, 0.687, 0.689, 0.692, 0.693, 0.695$, where the system exhibits shear jamming (from right to left in {\bf E}).
The yielding line (open brown diamonds) for the CV-S protocol are determined from the peak positions on the $\chi_\sigma - \gamma$ curves shown
in {\bf (F)}.
{\bf (G-I)} G-EOSs obtained by the {\color{black}CS-C/D}  protocol.
The solid (decompression) and dashed (compression) thin lines represent constant-$\gamma$ lines.  
The melting point (pink cross) is defined as the intersection between G-EOSs and L-EOS (think green line in {\bf G}).
The half filled brown diamonds are the yielding line for  the {\color{black}CS-C/D} protocol determined from
the peak positions on the $\chi_\sigma - \gamma$ curves shown in {\bf (I)}. See Fig.~\ref{fig:PD} for the meaning of other symbols.
}
\label{fig:EOS}
\end{figure*}

\subsection*{Shear-yielding and shear-jamming}

Up to now we have investigated
{\color{black} the interior of the stability-reversibility map.
Next} we turn to explore the boundaries of the stability-reversibility map by analyzing the G-EOSs both in pressure and shear stress. 
From now on, all data presented are averaged over different samples and realizations.
Therefore even with a finite size system the individual plastic events will be averaged out. Furthermore we will plot the G-EOSs on a phase diagram using $1/\varphi$ (instead of, equivalently,  $\epsilon$) and $\gamma$, in order to better show their relations to L-EOS.

First of all, starting from an equilibrium configuration at $(\varphi, \gamma) = (\varphi_{\rm g}, 0)$  or $(\epsilon,\gamma)=(0,0)$, the system melts under decompression
for sufficiently large $\epsilon$.
We define the melting point as the crossover point between the $\gamma=0$ G-EOS for the pressure and the L-EOS (see Fig.~\ref{fig:G-EOS}C and Fig.~\ref{fig:EOS}G). The melting point sets the upper bound of the stability-reversibility map along the $\gamma=0$ line.

To systematically explore the stability-reversibility map,
we design three specific protocols combining compression/decompression and shear, namely constant pressure-shear (CP-S), constant volume-shear (CV-S), and constant shear strain-compression/decompression ({\color{black}CS-C/D}); see Materials and Methods for details. These protocols can be realized also in experiments.
{\color{black} In principle the EOS should not be protocol-dependent, but whether it is also the case for G-EOS is not so obvious.}

In the CP-S protocol, for any fixed pressure $p$, the specific volume $1/\varphi$ (or {\color{black} volume strain}
$\epsilon$) evolves with shear strain $\gamma$, which defines a G-EOS for the pressure. 
Figure~\ref{fig:EOS}A shows the G-EOSs for a few different pressures $p$ in a $\gamma - 1/\varphi$ plot. Such a plot  is essentially 
the projection of the three-dimensional plot of the G-EOSs for the pressure $p=p(\varphi_{\rm g};\epsilon,\gamma)$
in Fig.~\ref{fig:G-EOS}C onto the $\gamma - 1/\varphi$  plane.
The data show that the specific volume $1/\varphi$ expands as strain $\gamma$ is increased, known as the dilatancy effect. 
{\color{black}
The dilatancy is stronger for better annealed glasses, as observed previously in Ref.~\cite{jin2017exploring},
and at lower pressure for a fixed quality of annealing, as shown here. Both observations are consistent with 
theoretical predictions (Fig.~2 in Ref.~\cite{RUYZ2015PRL} and Fig.~2a in Ref.~\cite{UZ17}, respectively).
Note that this dilatancy effect  shall be distinguished from the one discussed in the context of steady flow, which is necessarily correlated to friction as shown in~\cite{peyneau2008frictionless}.
}
At high pressures,  the isobaric lines are nearly parallel to the shear-jamming line,  which corresponds to the $p = \infty$ isobaric line by the definition of jamming.
On the other hand, the average stress $\sigma$ shown in Fig.~\ref{fig:EOS}B
initially increases with the shear strain, but it eventually approaches a plateau after a big drop corresponding to yielding. We define the yielding point as the peak of the stress susceptibility $\chi_\sigma = N (\langle \sigma^2 \rangle - \langle \sigma \rangle^2)$ (see Fig.~\ref{fig:EOS}C).
The yielding point is approximately at the middle of the drop on the stress-strain curve, corresponding to the steepest decrease of stress. After yielding, the shear stress generally remains non-zero, indicating that the glass is not completely fluidized. Indeed, real-space visualization shows that the glass breaks into two pieces sliding against each other (see Fig.~\ref{fig:PD}C).
However, near the melting point, such a picture might change, because melting could mix with yielding giving rise to a hybrid behavior. We will not discuss this situation in detail here.
Connecting the yielding strains obtained at different  $p$ we obtain the yielding line.
We notice that for a certain range of pressure $p$ near $p_{\rm g}$ of the initial glass,
the yielding strain $\gamma_{\rm Y}^* \approx 0.118$ is nearly independent of $p$.

In Fig.~\ref{fig:EOS}D and E, we show how the inverse pressure $1/p$ and shear stress $\sigma$ evolve with $\gamma$ for various $\varphi$, in the CV-S protocol. We find a threshold density $\varphi_{\rm c}$ (see Fig.~S6 for how  $\varphi_{\rm c}$ is determined ), which separates the shear yielding and shear jamming cases.
If $\varphi < \varphi_{\rm c}$, the system generally yields at large $\gamma$; otherwise both pressure and shear stress diverge as $\gamma$ is increased, indicating shear jamming.
In this protocol, the yielding point can be determined again from the peak of the stress susceptibility (see Fig.~\ref{fig:EOS}F).
In the shear jamming case, the pressure and shear stress both follow the free-volume scaling laws: $p \sim (\gamma_{\rm J}- \gamma)^{-1}$ and $\sigma \sim (\gamma_{\rm J}- \gamma)^{-1}$ (see Fig. S7). 
The shear jamming is a natural consequence of the dilatancy effect (i.e., $p$ increases with $\gamma$ for fixed $\varphi$), as long as the system does not yield. Thus $\varphi_{\rm c}$ results from the competition between the dilatancy effect and the tendency to break the system at large strains.
We have checked that {\color{black} all the shear jammed packings that we create satisfy the isostatic condition~\cite{maxwell1864calculation},} i.e., the average coordination number $Z=6$, once the ratters (particles who have less than four contacts) are excluded, and that the shear jamming transition falls in the same universality
class of the usual jamming transition in absence of shear.

Fig.~\ref{fig:EOS}G and \ref{fig:EOS}H show the constant-$\gamma$ EOSs of the pressure and shear stress
for a few different $\gamma$ in the {\color{black}CS-C/D} protocol.
For small shear strains,
$\gamma < \gamma_{\rm Y}^*$, the system jams at a $\gamma$-dependent jamming density $\varphi_{\rm J}$ under compression.
For shear strains larger than the yielding strain 
$\gamma > \gamma_{\rm Y}^*$, however, the G-EOSs for pressure collapse onto the same curve, and consequently,  the jamming density $\varphi_{\rm J}$ also does not change with $\gamma$ anymore. This observation is consistent with our interpretation of yielded states: the glass just breaks into two pieces of solids at $\gamma_{\rm Y}^*$ by forming a planar fracture. Such planar structures should have minor effect on bulk properties like the pressure.
  On the other hand, the glass always melts under decompression, for any $\gamma$.
We find that the melting point
is independent of $\gamma$, both below and above $\gamma_{\rm Y}^*$. Interestingly, 
the stress susceptibility $\chi_{\rm \sigma}$ displays a peak upon decompression, which reveals the vestige of yielding, and therefore can be used to define the yielding point in the {\color{black}CS-C/D} protocol (Fig.~\ref{fig:EOS}{\color{black}I}). For $\gamma <  \gamma_{\rm Y}^*$, the yielding density $\varphi$ increases with  $\gamma$; for $\gamma > \gamma_{\rm Y}^*$, the peak does not exist anymore and the yielding point cannot be defined as expected. In addition,  we show and discuss the behavior of the pressure susceptibility $\chi_p$ in Fig.~S8.

\begin{figure*}
\centerline{\includegraphics[width=0.9\columnwidth]{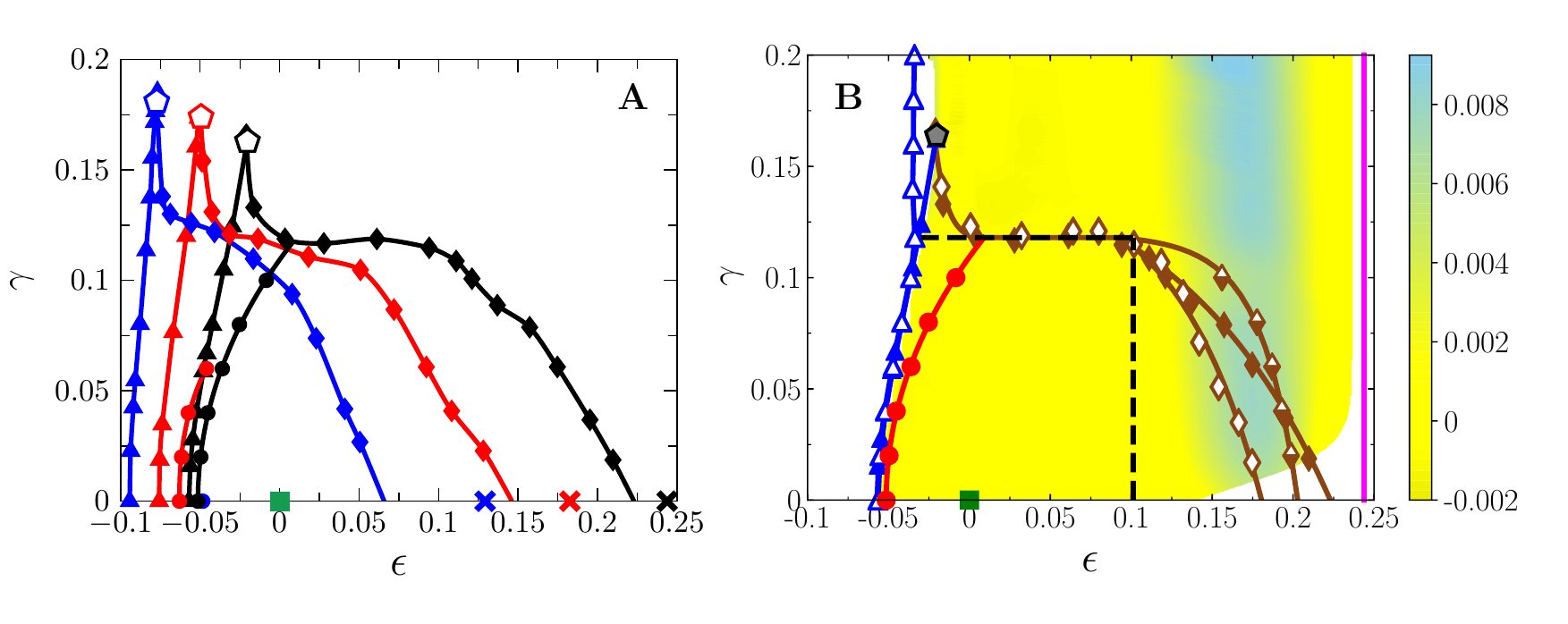} }
\caption{{\bf Protocol dependence of the stability-reversibility map.} {\bf (A)} The stability-reversibility maps for $\varphi_{\rm g} = 0.609$ (blue), 0.631 (red), 0.655 (black).  {\bf (B)} The stability-reversibility maps obtained in three different protocols (CP-S, CV-S and {\color{black}CS-C/D}). The color bar indicates  the difference on $1/p$ between the {\color{black}CS-C/D} and the CP-S protocols. The horizontal and vertical lines represent $\gamma_{\rm Y}^* \approx 0.118$ and $\epsilon^* \approx 0.1$. See Figs.~\ref{fig:PD} and~\ref{fig:EOS} for the meaning of symbols.}
\label{fig:protocol}
\end{figure*}

\subsection*{Dependence on protocols and system sizes}
Let us discuss how the stability-reversibility map and G-EOSs depend on protocols. There are two important sources of protocol dependences. Firstly, the stability-reversibility map
and  the G-EOSs depend on the glass transition point $\varphi_{\rm g}$, and $\varphi_{\rm g}$ itself 
depends on the protocol parameters such as the compression rate in a standard compression annealing protocol (here it is a function of where we stop swap moves).
Fig.~\ref{fig:protocol}A shows the stability-reversibility maps for three different $\varphi_{\rm g}$, corresponding to three typical experimental time scales as discussed previously (see {\color{black} Fig.~S9A} for the three-dimensional representations). They share common qualitative features in general. The stable regime expands with $\varphi_{\rm g}$, as one would naturally expect that more deeply annealed glasses should be more stable. {\color{black} Interestingly, the shear jamming line becomes more vertical with decreasing $\varphi_{\rm g}$. This trend is consistent with previous numerical observations which show that, in the thermodynamical limit,  the shear jamming line is completely vertical for infinitely rapidly quenched systems~\cite{baity2017emergent}.
Moreover,} we point out that the Gardner transition points can not be determined unambiguously using our approaches for the less annealed systems $\varphi_{\rm g} = 0.631, \gamma > 0.06$, and $\varphi_{\rm g} = 0.609, \gamma > 0$ (see Fig.~S5), because different activated dynamics, such as plastic rearrangements, formation of fractured structures, and $\alpha$-relaxations, cannot be well separated. 

Secondly, we show in Fig.~\ref{fig:protocol}B how the stability-reversibility map and also the G-EOSs for the pressure depends on the exploration protocols (CP-S, CV-S and {\color{black}CS-C/D}); see  {\color{black} Fig.~S9B and C} for the three-dimensional representations.
We find a protocol-independent regime $(\epsilon < \epsilon^{*},\gamma < \gamma_{\rm Y}^* )$, where all three protocols give the same pressure.
The part of the stability-reversibility map above $\gamma_{\rm Y}^*$ cannot be accessed by the {\color{black}CS-C/D} protocol. For $\epsilon > \epsilon^*$, the yielding line bends down differently depending on the protocol. The system yields most easily in the CV-S protocol, presumably because the liquid bubbles formed around melting are easier to expand in a volume-controlled protocol~\cite{fullerton2017density}. 


Finally we discuss briefly how the stability-reversibility maps depend on the system size $N$ in {\color{black} Fig.~S10}. 
We do not observe appreciable finite-size effects on the shear-jamming line.
{\color{black} 
On the other hand, the yielding line exhibits strong finite size effects, but we expect it to converge at larger sizes, based on the recent results of Ref.~\cite{ozawa2018random}.
}
Using the present methods, we also do not find strong $N$-dependence on the Gardner line, {\color{black} consistent} with the data in Ref.~\cite{jin2017exploring}. However, we stress that based on available numerical results, we cannot conclude on the thermodynamic behavior of the Gardner transition. 
Understanding whether it is a sharp transition or a crossover is an active and hot topic in the field, through
numerical~\cite{BCJPSZ2016PNAS}, 
experimental~\cite{seguin2016experimental} and 
theoretical analysis~\cite{BU15, charbonneau2017nontrivial, lubchenko2017aging, hicks2017gardner}. 
{\color{black} While the finite-size analysis presented here shall not be considered as conclusive, 
we leave a more detailed finite-size study on yielding,
shear jamming, and the Gardner transition for future works.}

\section*{Discussion}

\FZ{In this paper, we investigate the stability and the reversibility of polydisperse hard sphere glasses under volume and shear strains. We prepare equilibrium supercooled liquid states, with different degrees of 
stability ranging from a fast quench to a extremely slow annealing, corresponding to ultra-stable configurations.
Each configuration corresponds to a glass within a time scale that is shorter than the structural relaxation time. We study the stability of the glass under volume and shear strains, and find that the region of stability is delimited
by lines where the system can either yield or jam. We also find that within the region of stability, the system can be either a normal solid which essentially responds elastically and
reversibly to perturbations, or a marginally stable solid, which responds plastically and in a partially irreversible way.
More precisely, the main outcome of our analysis is the following:}
\begin{enumerate}

\item
{\bf Response.}
The response of the system to a shear strain is either purely elastic, partially plastic, or
  fully plastic (yielding),
  depending on the quality of annealing and the amount of {\color{black} volume} and shear strains 
  imposed to it.
  
\item
{\bf Failure.}
Well annealed glasses (large $\varphi_{\rm g}$), when sheared at sufficiently low densities (large {\color{black} volume strain} $\epsilon$), behave purely
  elastically up to yielding, which is
 an abrupt {\color{black} process}
 where a fracture is formed and the glass fails.
  At higher densities, they display a partial plastic phase before yielding is reached.
  At  even higher densities, they display shear-jamming (under constant volume shear).
  The shear-yielding and shear-jamming lines delimit the region of existence of the HS amorphous solid.

\item
{\bf Marginality.}
Along the solid part of the stress-strain curves, the partial plastic behavior is well separated from the purely elastic one by the Gardner point.
The onset of partial plasticity is accompanied by the emergence of 
critical behavior and 
marginal stability. Beyond the Gardner point, 
the shear modulus of the system becomes history dependent. 
At the same time, a growing {\color{black} spin-glass--like} susceptibility is observed.

\item
{\bf Reversibility.}
The purely elastic phase is globally reversible: once the shear is released, the system gets back to the original configuration.
The partially plastic marginal glass phase is partially irreversible: upon releasing the deformation by a small amount,
 the system is not able to get back to the previous state, while  upon complete release, the system is able to get back to the original configuration.
 Yielding corresponds  to complete irreversibility: once broken, the system starts to flow and it is not able to get back to the original configuration once the strain is completely released.

\end{enumerate}

Collecting together the boundaries of the different regions, we obtained a complete stability-reversibility map (phase diagram),
reported in Fig.~\ref{fig:PD}.
The stability-reversibility map obtained in the present study for three dimensional HS glasses
can be compared with the  one obtained by the MF theory in the large dimensional limit \cite{UZ17}.
The most important features, such as the
presence of the shear jamming and the shear yielding lines that delimit the stability region,
and the presence of the Gardner line, are qualitatively in good agreement with the predictions of the
 theory. There are, however, several important differences. 
(i)~The
shear-yielding line in the three dimensional system is not a spinodal line as predicted by the MF theory~\cite{RUYZ2015PRL}. The abrupt formation of a fracture is completely missed by the MF theory,
 which does not describe the spatial fluctuations of stress that accumulate around the fracture.
(ii)~The point $(\epsilon_{\rm c}, \gamma_{\rm c})$ where the shear-yielding and the shear-jamming lines meet is predicted 
to be a critical point in the MF theory, but it is rather a crossover point in the three-dimensional
system. (iii)~Quite interestingly, the marginally stable phase has larger $\gamma_{\rm Y}$
than the stable phase. 
This suggests that the plastic events in the marginal phase
help the system to avoid total failure. In the theory, the shear-yielding line bends down rather than bends up near the point $(\epsilon_{\rm c}, \gamma_{\rm c})$ (see Fig. 3 of Ref.~\cite{UZ17}).

Note that  the MF predictions of Ref.~\cite{UZ17} were obtained using the so-called replica symmetric (RS) ansatz. 
To properly consider yielding in the marginally stable phase, one should extend the computation to a full-step 
replica symmetric breaking (fullRSB) ansatz~\cite{CP2016}. 
This might help solving some of the discrepancies between the analytical and the numerical results.
According to the RS theory, yielding is a spinodal transition with disorder
~\cite{RUYZ2015PRL}. 
However, it is not clear how this picture will be modified by a fullRSB theory.

Our simulation results show that a well-annealed glass ($\varphi_{\rm g}$ well above the MCT density  $\varphi_{\rm MCT}$) yields abruptly -- it is brittle. 
However, a poorly annealed glass ($\varphi_{\rm g} \sim \varphi_{\rm MCT}$) may instead continuously yield
into a plastic flow state~\cite{bonn2017yield, AFMB17} -- it is more ductile. 
We expect that near the melting point, even a well-annealed glass would behave similarly to a poorly annealed one, as it would become much ``softer" upon decompression. Nevertheless, the yielding point can be  determined from the peak of $\chi_{\sigma}$ for both cases as shown here. Our approach thus provides a unified framework to study the transition between the two distinct mechanisms of yielding. The possibility of two yielding mechanisms is
missed by the current 
MF theory.
A dynamical extension of MF might account for such effects. Understanding the nature of the yielding transition~\cite{JPRS16, PPRS17,wisitsorasak2012strength} is a crucial problem which requires further analysis.

The plastic events we observe in the partially irreversible phase 
could correspond to two different types of soft modes: collective modes, associated to a diverging length scale,
 as predicted by the MF theory in the marginally stable phase~\cite{muller2015marginal,CKPUZ17}; or localized modes, such as the ones that
 have been observed in numerical studies  of low-dimensional systems~\cite{ML11, Mizuno14112017, lerner2016statistics}. 
 In this study, we did not investigate systematically the nature of the plastic events in our system, but 
 the growth of the {\color{black} spin-glass--like} susceptibility in our data
 suggests the presence, in our HS model, 
 of large-scale collective excitations.
Note that the situation could be radically different in 
soft-potential models~\cite{Mizuno14112017, lerner2016statistics,PhysRevLett.119.205501}. We would also like to stress that while the existence of partial plasticity before yielding is well-known~\cite{schuh2007mechanical,rodney2011modeling,barrat2011heterogeneities}, our well-annealed systems provide an example
where the pure elasticity and partial plasticity regimes are well separated, allowing us to define a line (the Gardner line) that separates them in the stability-reversibility map.

Finally, concerning the reversibility,
here we focus on the reversibility with respect to just one cycle of simple shear (see {\color{black} Fig.~S2} for the results under a few cycles). 
In cyclic shear protocols, a steady state can be reached after many cycles \cite{fiocco2013oscillatory,kawasaki2016macroscopic}. Very complicated dynamics should be involved in such processes. It would be interesting to systematically extend the present study to multiple cyclic shear, in order to understand better such processes.

\section*{Materials and Methods}

\subsection*{Model}
The system consists of $N=1000$ (unless otherwise specified) HS particles with a diameter distribution $P(D) \sim D^{-3}$, for $D_{\rm min} \leq D \leq D_{\rm min}/0.45$.  The continuous polydispersity is sufficient to suppress crystallization even in deep annealing, and optimizes the efficiency  of swap algorithm. 
\YJ{The volume fraction is
$\varphi=\rho (4/3)\pi \overline{D^3}$,
where $\rho=N/V$ is the number density and $V$ is the total volume.}
We define the reduced pressure $p=PV/Nk_{\rm B}T$ and the reduced stress $\sigma=\Sigma V/Nk_{\rm B}T$, where $P$ and $\Sigma$ are the  pressure and the stress of the system. 
For simplicity, in the rest of this paper we refer as pressure and stress to $p$ and $\sigma$ instead of $P$ and $\Sigma$.
We set Boltzmann constant $k_{\rm B}$, the temperature $T$, the mean diameter $\overline{D}$, and the particle mass $m$ to unity.  

\subsection*{Swap algorithm}
 At each dynamical step, the swap Monte Carlo algorithm attempts to exchange the positions of two randomly picked particles as long as they do not overlap  with their new neighbors. Such non-local Monte Carlo moves  eliminate the local confinement of particles in supercooled states, 
 which, combined with standard event-driven MD, significantly facilitates the equilibration procedure. It has been carefully examined that the swap algorithm does not introduce crystalline order in the polydisperse HS model studied here~\cite{BCNO2016PRL}.

\subsection*{Compression/decompression algorithm}
We use the Lubachevsky-Stillinger algorithm~\cite{lubachevsky1990geometric} to compress and decompress the system. The particles are simulated by using event-driven MD. 
The sphere diameters are increased/decreased with a constant rate.
The MD time is expressed in units of $\sqrt{1/k_{\rm B} m\overline{D}^2}$. 

\subsection*{Simple shear algorithm}
At each step, we perform $N_{\rm collision} = 100 -1000$ collisions per particle using the event-driven MD, and then instantaneously increase the shear strain by $\delta \gamma = \dot{\gamma} \delta t$, where  $\delta t$ is the time elapsed during the collisions. The instantaneous shear shifts all particles by $x_i \to x_i + \delta \gamma z_i$, where $x_i$ and $z_i$ are the $x-$ and $z-$coordinates of particle $i$. To remove the possible 
overlappings introduced during this shift, we switch to a harmonic inter-particle potential and use the conjugated gradient (CG) method to minimize the energy. The harmonic potential is switched off after CG. The Lees-Edwards boundary conditions~\cite{lees1972computer} are used. 
See Ref.~\cite{jin2017exploring} for more details.

\subsection*{Protocols of zero-field compression (ZFC) and field compression (FC)}
In the ZFC protocol, starting from the initial equilibrium configuration at  $(\epsilon, \gamma) = (0,0)$, 
we (i) firstly shear the system to a target shear strain at $(0, \gamma)$
 while keeping the {\color{black} volume strain} unchanged, (ii) secondly compress it to a target {\color{black} volume strain} at $( \epsilon, \gamma )$ keeping the shear strain unchanged,  (iii) then apply an additional small shear strain $\delta \gamma = 0.002$, (iv) and finally measure the stress $\sigma_{\rm ZFC}$ at the state point $( \epsilon, \gamma +\delta \gamma )$. In the FC protocol, the order of steps (ii) and (iii) are interchanged. The FC protocol therefore has the path $(0,0) \to (0, \gamma) \to (0, \gamma + \delta \gamma )  \to ( \epsilon, \gamma + \delta \gamma ) $. The target shear strain is chosen such that it is below the yielding strain $\gamma < \gamma_{\rm Y}$.  Here the shear strain serves as an external ``field" with respect to compression, in analogy to the magnetic field in  cooling experiments on spin glasses~\cite{nagata1979low}. 
 {\color{black} The stress is measured on a time scale $t = 10 \approx 10\tau_0$, where $\tau_0$ is the ballistic time. This choice ensures that the ZFC protocol measures the short time response to shear, while the FC measurement corresponds to the long time response because the shear strain $\gamma +\delta \gamma$ is
reached before the volume strain is applied
   (see Ref.~\cite{jin2017exploring} for a detailed analysis on the stress relaxation dynamics.)}
 This protocol generalizes the one used in Ref.~\cite{jin2017exploring} which corresponds to the case $\gamma=0$. 

\subsection*{Protocols of constant pressure-shear (CP-S), constant volume-shear (CV-S), and constant  shear strain-compression/decompression ({\color{black}CS-C/D})}
In the CP-S protocol, the system is firstly compressed or decompressed (depending on if the target $p$ is higher or lower than $p_{\rm g}$) from the equilibrium state at $(p, \gamma) = (p_{\rm g}, 0)$ to the  state at $(p, 0)$. Then simple shear is applied under the constant-$p$ condition, until the system reaches the target shear strain at $(p, \gamma)$. At each shear step, the particle diameters are adjusted to keep $p$ constant. In the CV-S protocol, the system is firstly compressed or decompressed from $\varphi= \varphi_{\rm g}$ to the target density $\varphi$, and then the simple shear is applied by keeping the volume constant. In the {\color{black}CS-C/D} protocol, the system is firstly sheared from $(\varphi, \gamma) = (\varphi_{\rm g}, 0)$ to a target strain at  $(\varphi_{\rm g}, \gamma)$, and then compression or decompression is applied while keeping the shear strain $\gamma$ constant. 

\subsection*{Caging order parameters}
We consider two order parameters {\color{black} $\Delta_{\rm r}$ and $\Delta_{AB}$ defined below to characterize the glass state}.
The relative mean squared displacement is defined as
\beq
\Delta_{\rm r } = \frac{1}{N} \sum_{i=1}^{N}   \left |\vr_i  - \vr_i^{\rm r} \right |^2,   
 \eeq
 where $\{ \vr_i \}$ and $\{\vr_i^{\rm r}\}$ are the particle coordinates of the target and reference configurations. In Fig.~\ref{fig:reversibility}, the target and reference are the configurations after and before shear respectively. The replica mean squared displacement 
 \beq
 \Delta_{AB} = \frac{1}{N} \sum_{i=1}^{N}   \left |\vr_i^A  - \vr_i^B \right |^2,   
 \eeq
measures the distance between two replicas of the same sample  generated by two independent realizations. 
  
One may also consider the time-dependent mean squared displacement $\Delta(t) = \frac{1}{N} \sum_{i=1}^{N} \langle  |\vr_i(t)  - \vr_i(0) |^2  \rangle$, whose value at the ballistic time scale $\tau_0 \sim 1$ gives the typical vibrational cage size of particles. We found that in our systems,  $\Delta(\tau_0) \lesssim 0.01$, see Ref.~\cite{BCJPSZ2016PNAS}. The cage size is nearly unchanged under simple shear.


\bibliography{HSshear,yoshino}

\begin{thebibliography}{59}%
\makeatletter
\providecommand \@ifxundefined [1]{%
 \@ifx{#1\undefined}
}%
\providecommand \@ifnum [1]{%
 \ifnum #1\expandafter \@firstoftwo
 \else \expandafter \@secondoftwo
 \fi
}%
\providecommand \@ifx [1]{%
 \ifx #1\expandafter \@firstoftwo
 \else \expandafter \@secondoftwo
 \fi
}%
\providecommand \natexlab [1]{#1}%
\providecommand \enquote  [1]{``#1''}%
\providecommand \bibnamefont  [1]{#1}%
\providecommand \bibfnamefont [1]{#1}%
\providecommand \citenamefont [1]{#1}%
\providecommand \href@noop [0]{\@secondoftwo}%
\providecommand \href [0]{\begingroup \@sanitize@url \@href}%
\providecommand \@href[1]{\@@startlink{#1}\@@href}%
\providecommand \@@href[1]{\endgroup#1\@@endlink}%
\providecommand \@sanitize@url [0]{\catcode `\\12\catcode `\$12\catcode
  `\&12\catcode `\#12\catcode `\^12\catcode `\_12\catcode `\%12\relax}%
\providecommand \@@startlink[1]{}%
\providecommand \@@endlink[0]{}%
\providecommand \url  [0]{\begingroup\@sanitize@url \@url }%
\providecommand \@url [1]{\endgroup\@href {#1}{\urlprefix }}%
\providecommand \urlprefix  [0]{URL }%
\providecommand \Eprint [0]{\href }%
\providecommand \doibase [0]{http://dx.doi.org/}%
\providecommand \selectlanguage [0]{\@gobble}%
\providecommand \bibinfo  [0]{\@secondoftwo}%
\providecommand \bibfield  [0]{\@secondoftwo}%
\providecommand \translation [1]{[#1]}%
\providecommand \BibitemOpen [0]{}%
\providecommand \bibitemStop [0]{}%
\providecommand \bibitemNoStop [0]{.\EOS\space}%
\providecommand \EOS [0]{\spacefactor3000\relax}%
\providecommand \BibitemShut  [1]{\csname bibitem#1\endcsname}%
\let\auto@bib@innerbib\@empty
\bibitem [{\citenamefont {Wang}\ \emph {et~al.}(2004)\citenamefont {Wang},
  \citenamefont {Dong},\ and\ \citenamefont {Shek}}]{WDS04}%
  \BibitemOpen
  \bibfield  {author} {\bibinfo {author} {\bibfnamefont {Wei-Hua}\ \bibnamefont
  {Wang}}, \bibinfo {author} {\bibfnamefont {Chuang}\ \bibnamefont {Dong}}, \
  and\ \bibinfo {author} {\bibfnamefont {CH}~\bibnamefont {Shek}},\ }\bibfield
  {title} {\enquote {\bibinfo {title} {Bulk metallic glasses},}\ }\href@noop {}
  {\bibfield  {journal} {\bibinfo  {journal} {Materials Science and
  Engineering: R: Reports}\ }\textbf {\bibinfo {volume} {44}},\ \bibinfo
  {pages} {45--89} (\bibinfo {year} {2004})}\BibitemShut {NoStop}%
\bibitem [{\citenamefont {Angell}\ \emph {et~al.}(2000)\citenamefont {Angell},
  \citenamefont {Ngai}, \citenamefont {McKenna}, \citenamefont {McMillan},\
  and\ \citenamefont {Martin}}]{angell2000relaxation}%
  \BibitemOpen
  \bibfield  {author} {\bibinfo {author} {\bibfnamefont {C~Austin}\
  \bibnamefont {Angell}}, \bibinfo {author} {\bibfnamefont {Kia~L}\
  \bibnamefont {Ngai}}, \bibinfo {author} {\bibfnamefont {Greg~B}\ \bibnamefont
  {McKenna}}, \bibinfo {author} {\bibfnamefont {Paul~F}\ \bibnamefont
  {McMillan}}, \ and\ \bibinfo {author} {\bibfnamefont {Steve~W}\ \bibnamefont
  {Martin}},\ }\bibfield  {title} {\enquote {\bibinfo {title} {Relaxation in
  glassforming liquids and amorphous solids},}\ }\href@noop {} {\bibfield
  {journal} {\bibinfo  {journal} {Journal of Applied Physics}\ }\textbf
  {\bibinfo {volume} {88}},\ \bibinfo {pages} {3113--3157} (\bibinfo {year}
  {2000})}\BibitemShut {NoStop}%
\bibitem [{\citenamefont {Cavagna}(2009)}]{Ca09}%
  \BibitemOpen
  \bibfield  {author} {\bibinfo {author} {\bibfnamefont {Andrea}\ \bibnamefont
  {Cavagna}},\ }\bibfield  {title} {\enquote {\bibinfo {title} {Supercooled
  liquids for pedestrians},}\ }\href@noop {} {\bibfield  {journal} {\bibinfo
  {journal} {Physics Reports}\ }\textbf {\bibinfo {volume} {476}},\ \bibinfo
  {pages} {51--124} (\bibinfo {year} {2009})}\BibitemShut {NoStop}%
\bibitem [{\citenamefont {Liu}\ and\ \citenamefont {Nagel}(2010)}]{LN10}%
  \BibitemOpen
  \bibfield  {author} {\bibinfo {author} {\bibfnamefont {Andrea~J}\
  \bibnamefont {Liu}}\ and\ \bibinfo {author} {\bibfnamefont {Sidney~R}\
  \bibnamefont {Nagel}},\ }\bibfield  {title} {\enquote {\bibinfo {title} {The
  jamming transition and the marginally jammed solid},}\ }\href@noop {}
  {\bibfield  {journal} {\bibinfo  {journal} {Annu. Rev. Condens. Matter
  Phys.}\ }\textbf {\bibinfo {volume} {1}},\ \bibinfo {pages} {347--369}
  (\bibinfo {year} {2010})}\BibitemShut {NoStop}%
\bibitem [{\citenamefont {Charbonneau}\ \emph {et~al.}(2017)\citenamefont
  {Charbonneau}, \citenamefont {Kurchan}, \citenamefont {Parisi}, \citenamefont
  {Urbani},\ and\ \citenamefont {Zamponi}}]{CKPUZ17}%
  \BibitemOpen
  \bibfield  {author} {\bibinfo {author} {\bibfnamefont {Patrick}\ \bibnamefont
  {Charbonneau}}, \bibinfo {author} {\bibfnamefont {Jorge}\ \bibnamefont
  {Kurchan}}, \bibinfo {author} {\bibfnamefont {Giorgio}\ \bibnamefont
  {Parisi}}, \bibinfo {author} {\bibfnamefont {Pierfrancesco}\ \bibnamefont
  {Urbani}}, \ and\ \bibinfo {author} {\bibfnamefont {Francesco}\ \bibnamefont
  {Zamponi}},\ }\bibfield  {title} {\enquote {\bibinfo {title} {Glass and
  jamming transitions: From exact results to finite-dimensional
  descriptions},}\ }\href@noop {} {\bibfield  {journal} {\bibinfo  {journal}
  {Annual Review of Condensed Matter Physics}\ }\textbf {\bibinfo {volume}
  {8}},\ \bibinfo {pages} {265--288} (\bibinfo {year} {2017})}\BibitemShut
  {NoStop}%
\bibitem [{\citenamefont {Schuh}\ \emph {et~al.}(2007)\citenamefont {Schuh},
  \citenamefont {Hufnagel},\ and\ \citenamefont
  {Ramamurty}}]{schuh2007mechanical}%
  \BibitemOpen
  \bibfield  {author} {\bibinfo {author} {\bibfnamefont {Christopher~A}\
  \bibnamefont {Schuh}}, \bibinfo {author} {\bibfnamefont {Todd~C}\
  \bibnamefont {Hufnagel}}, \ and\ \bibinfo {author} {\bibfnamefont
  {Upadrasta}\ \bibnamefont {Ramamurty}},\ }\bibfield  {title} {\enquote
  {\bibinfo {title} {Mechanical behavior of amorphous alloys},}\ }\href@noop {}
  {\bibfield  {journal} {\bibinfo  {journal} {Acta Materialia}\ }\textbf
  {\bibinfo {volume} {55}},\ \bibinfo {pages} {4067--4109} (\bibinfo {year}
  {2007})}\BibitemShut {NoStop}%
\bibitem [{\citenamefont {Rodney}\ \emph {et~al.}(2011)\citenamefont {Rodney},
  \citenamefont {Tanguy},\ and\ \citenamefont
  {Vandembroucq}}]{rodney2011modeling}%
  \BibitemOpen
  \bibfield  {author} {\bibinfo {author} {\bibfnamefont {David}\ \bibnamefont
  {Rodney}}, \bibinfo {author} {\bibfnamefont {Anne}\ \bibnamefont {Tanguy}}, \
  and\ \bibinfo {author} {\bibfnamefont {Damien}\ \bibnamefont
  {Vandembroucq}},\ }\bibfield  {title} {\enquote {\bibinfo {title} {Modeling
  the mechanics of amorphous solids at different length scale and time
  scale},}\ }\href@noop {} {\bibfield  {journal} {\bibinfo  {journal} {Model.
  Simul. Mater. Sci. Eng.}\ }\textbf {\bibinfo {volume} {19}},\ \bibinfo
  {pages} {083001} (\bibinfo {year} {2011})}\BibitemShut {NoStop}%
\bibitem [{\citenamefont {Barrat}\ and\ \citenamefont
  {Lemaˆ{\i}tre}(2011)}]{barrat2011heterogeneities}%
  \BibitemOpen
  \bibfield  {author} {\bibinfo {author} {\bibfnamefont {Jean-Louis}\
  \bibnamefont {Barrat}}\ and\ \bibinfo {author} {\bibfnamefont {Ana{\"e}l}\
  \bibnamefont {Lemaˆ{\i}tre}},\ }\bibfield  {title} {\enquote {\bibinfo
  {title} {Heterogeneities in amorphous systems under shear},}\ }\href@noop {}
  {\bibfield  {journal} {\bibinfo  {journal} {Dynamical heterogeneities in
  glasses, colloids, and granular media}\ }\textbf {\bibinfo {volume} {150}},\
  \bibinfo {pages} {264} (\bibinfo {year} {2011})}\BibitemShut {NoStop}%
\bibitem [{\citenamefont {Ozawa}\ \emph {et~al.}(2018)\citenamefont {Ozawa},
  \citenamefont {Berthier}, \citenamefont {Biroli}, \citenamefont {Rosso},\
  and\ \citenamefont {Tarjus}}]{ozawa2018random}%
  \BibitemOpen
  \bibfield  {author} {\bibinfo {author} {\bibfnamefont {Misaki}\ \bibnamefont
  {Ozawa}}, \bibinfo {author} {\bibfnamefont {Ludovic}\ \bibnamefont
  {Berthier}}, \bibinfo {author} {\bibfnamefont {Giulio}\ \bibnamefont
  {Biroli}}, \bibinfo {author} {\bibfnamefont {Alberto}\ \bibnamefont {Rosso}},
  \ and\ \bibinfo {author} {\bibfnamefont {Gilles}\ \bibnamefont {Tarjus}},\
  }\bibfield  {title} {\enquote {\bibinfo {title} {A random critical point
  separates brittle and ductile yielding transitions in amorphous materials},}\
  }\href@noop {} {\bibfield  {journal} {\bibinfo  {journal} {arXiv:1803.11502}\
  } (\bibinfo {year} {2018})}\BibitemShut {NoStop}%
\bibitem [{\citenamefont {Bonn}\ \emph {et~al.}(2017)\citenamefont {Bonn},
  \citenamefont {Denn}, \citenamefont {Berthier}, \citenamefont {Divoux},\ and\
  \citenamefont {Manneville}}]{bonn2017yield}%
  \BibitemOpen
  \bibfield  {author} {\bibinfo {author} {\bibfnamefont {Daniel}\ \bibnamefont
  {Bonn}}, \bibinfo {author} {\bibfnamefont {Morton~M}\ \bibnamefont {Denn}},
  \bibinfo {author} {\bibfnamefont {Ludovic}\ \bibnamefont {Berthier}},
  \bibinfo {author} {\bibfnamefont {Thibaut}\ \bibnamefont {Divoux}}, \ and\
  \bibinfo {author} {\bibfnamefont {S{\'e}bastien}\ \bibnamefont
  {Manneville}},\ }\bibfield  {title} {\enquote {\bibinfo {title} {Yield stress
  materials in soft condensed matter},}\ }\href@noop {} {\bibfield  {journal}
  {\bibinfo  {journal} {Reviews of Modern Physics}\ }\textbf {\bibinfo {volume}
  {89}},\ \bibinfo {pages} {035005} (\bibinfo {year} {2017})}\BibitemShut
  {NoStop}%
\bibitem [{\citenamefont {M{\"u}ller}\ and\ \citenamefont
  {Wyart}(2015)}]{muller2015marginal}%
  \BibitemOpen
  \bibfield  {author} {\bibinfo {author} {\bibfnamefont {Markus}\ \bibnamefont
  {M{\"u}ller}}\ and\ \bibinfo {author} {\bibfnamefont {Matthieu}\ \bibnamefont
  {Wyart}},\ }\bibfield  {title} {\enquote {\bibinfo {title} {Marginal
  stability in structural, spin, and electron glasses},}\ }\href@noop {}
  {\bibfield  {journal} {\bibinfo  {journal} {Annual Review of Condensed Matter
  Physics}\ }\textbf {\bibinfo {volume} {6}},\ \bibinfo {pages} {177} (\bibinfo
  {year} {2015})}\BibitemShut {NoStop}%
\bibitem [{\citenamefont {Nicolas}\ \emph {et~al.}(2017)\citenamefont
  {Nicolas}, \citenamefont {Ferrero}, \citenamefont {Martens},\ and\
  \citenamefont {Barrat}}]{AFMB17}%
  \BibitemOpen
  \bibfield  {author} {\bibinfo {author} {\bibfnamefont {Alexandre}\
  \bibnamefont {Nicolas}}, \bibinfo {author} {\bibfnamefont {Ezequiel~E}\
  \bibnamefont {Ferrero}}, \bibinfo {author} {\bibfnamefont {Kirsten}\
  \bibnamefont {Martens}}, \ and\ \bibinfo {author} {\bibfnamefont
  {Jean-Louis}\ \bibnamefont {Barrat}},\ }\bibfield  {title} {\enquote
  {\bibinfo {title} {Deformation and flow of amorphous solids: a review of
  mesoscale elastoplastic models},}\ }\href@noop {} {\bibfield  {journal}
  {\bibinfo  {journal} {arXiv preprint arXiv:1708.09194}\ } (\bibinfo {year}
  {2017})}\BibitemShut {NoStop}%
\bibitem [{\citenamefont {Bi}\ \emph {et~al.}(2011)\citenamefont {Bi},
  \citenamefont {Zhang}, \citenamefont {Chakraborty},\ and\ \citenamefont
  {Behringer}}]{bi2011jamming}%
  \BibitemOpen
  \bibfield  {author} {\bibinfo {author} {\bibfnamefont {Dapeng}\ \bibnamefont
  {Bi}}, \bibinfo {author} {\bibfnamefont {Jie}\ \bibnamefont {Zhang}},
  \bibinfo {author} {\bibfnamefont {Bulbul}\ \bibnamefont {Chakraborty}}, \
  and\ \bibinfo {author} {\bibfnamefont {Robert~P}\ \bibnamefont {Behringer}},\
  }\bibfield  {title} {\enquote {\bibinfo {title} {Jamming by shear},}\
  }\href@noop {} {\bibfield  {journal} {\bibinfo  {journal} {Nature}\ }\textbf
  {\bibinfo {volume} {480}},\ \bibinfo {pages} {355} (\bibinfo {year}
  {2011})}\BibitemShut {NoStop}%
\bibitem [{\citenamefont {Peters}\ \emph {et~al.}(2016)\citenamefont {Peters},
  \citenamefont {Majumdar},\ and\ \citenamefont {Jaeger}}]{peters2016direct}%
  \BibitemOpen
  \bibfield  {author} {\bibinfo {author} {\bibfnamefont {Ivo~R}\ \bibnamefont
  {Peters}}, \bibinfo {author} {\bibfnamefont {Sayantan}\ \bibnamefont
  {Majumdar}}, \ and\ \bibinfo {author} {\bibfnamefont {Heinrich~M}\
  \bibnamefont {Jaeger}},\ }\bibfield  {title} {\enquote {\bibinfo {title}
  {Direct observation of dynamic shear jamming in dense suspensions},}\
  }\href@noop {} {\bibfield  {journal} {\bibinfo  {journal} {Nature}\ }\textbf
  {\bibinfo {volume} {532}},\ \bibinfo {pages} {214} (\bibinfo {year}
  {2016})}\BibitemShut {NoStop}%
\bibitem [{\citenamefont {Hyun}\ \emph {et~al.}(2011)\citenamefont {Hyun},
  \citenamefont {Wilhelm}, \citenamefont {Klein}, \citenamefont {Cho},
  \citenamefont {Nam}, \citenamefont {Ahn}, \citenamefont {Lee}, \citenamefont
  {Ewoldt},\ and\ \citenamefont {McKinley}}]{hyun2011review}%
  \BibitemOpen
  \bibfield  {author} {\bibinfo {author} {\bibfnamefont {Kyu}\ \bibnamefont
  {Hyun}}, \bibinfo {author} {\bibfnamefont {Manfred}\ \bibnamefont {Wilhelm}},
  \bibinfo {author} {\bibfnamefont {Christopher~O}\ \bibnamefont {Klein}},
  \bibinfo {author} {\bibfnamefont {Kwang~Soo}\ \bibnamefont {Cho}}, \bibinfo
  {author} {\bibfnamefont {Jung~Gun}\ \bibnamefont {Nam}}, \bibinfo {author}
  {\bibfnamefont {Kyung~Hyun}\ \bibnamefont {Ahn}}, \bibinfo {author}
  {\bibfnamefont {Seung~Jong}\ \bibnamefont {Lee}}, \bibinfo {author}
  {\bibfnamefont {Randy~H}\ \bibnamefont {Ewoldt}}, \ and\ \bibinfo {author}
  {\bibfnamefont {Gareth~H}\ \bibnamefont {McKinley}},\ }\bibfield  {title}
  {\enquote {\bibinfo {title} {A review of nonlinear oscillatory shear tests:
  Analysis and application of large amplitude oscillatory shear (laos)},}\
  }\href@noop {} {\bibfield  {journal} {\bibinfo  {journal} {Progress in
  Polymer Science}\ }\textbf {\bibinfo {volume} {36}},\ \bibinfo {pages}
  {1697--1753} (\bibinfo {year} {2011})}\BibitemShut {NoStop}%
\bibitem [{\citenamefont {Regev}\ \emph {et~al.}(2015)\citenamefont {Regev},
  \citenamefont {Weber}, \citenamefont {Reichhardt}, \citenamefont {Dahmen},\
  and\ \citenamefont {Lookman}}]{regev2015reversibility}%
  \BibitemOpen
  \bibfield  {author} {\bibinfo {author} {\bibfnamefont {Ido}\ \bibnamefont
  {Regev}}, \bibinfo {author} {\bibfnamefont {John}\ \bibnamefont {Weber}},
  \bibinfo {author} {\bibfnamefont {Charles}\ \bibnamefont {Reichhardt}},
  \bibinfo {author} {\bibfnamefont {Karin~A}\ \bibnamefont {Dahmen}}, \ and\
  \bibinfo {author} {\bibfnamefont {Turab}\ \bibnamefont {Lookman}},\
  }\bibfield  {title} {\enquote {\bibinfo {title} {Reversibility and
  criticality in amorphous solids},}\ }\href@noop {} {\bibfield  {journal}
  {\bibinfo  {journal} {Nature communications}\ }\textbf {\bibinfo {volume}
  {6}},\ \bibinfo {pages} {8805} (\bibinfo {year} {2015})}\BibitemShut
  {NoStop}%
\bibitem [{\citenamefont {Kawasaki}\ and\ \citenamefont
  {Berthier}(2016)}]{kawasaki2016macroscopic}%
  \BibitemOpen
  \bibfield  {author} {\bibinfo {author} {\bibfnamefont {Takeshi}\ \bibnamefont
  {Kawasaki}}\ and\ \bibinfo {author} {\bibfnamefont {Ludovic}\ \bibnamefont
  {Berthier}},\ }\bibfield  {title} {\enquote {\bibinfo {title} {Macroscopic
  yielding in jammed solids is accompanied by a nonequilibrium first-order
  transition in particle trajectories},}\ }\href@noop {} {\bibfield  {journal}
  {\bibinfo  {journal} {Physical Review E}\ }\textbf {\bibinfo {volume} {94}},\
  \bibinfo {pages} {022615} (\bibinfo {year} {2016})}\BibitemShut {NoStop}%
\bibitem [{\citenamefont {Leishangthem}\ \emph {et~al.}(2017)\citenamefont
  {Leishangthem}, \citenamefont {Parmar},\ and\ \citenamefont
  {Sastry}}]{leishangthem2017yielding}%
  \BibitemOpen
  \bibfield  {author} {\bibinfo {author} {\bibfnamefont {Premkumar}\
  \bibnamefont {Leishangthem}}, \bibinfo {author} {\bibfnamefont {Anshul~DS}\
  \bibnamefont {Parmar}}, \ and\ \bibinfo {author} {\bibfnamefont {Srikanth}\
  \bibnamefont {Sastry}},\ }\bibfield  {title} {\enquote {\bibinfo {title} {The
  yielding transition in amorphous solids under oscillatory shear
  deformation},}\ }\href@noop {} {\bibfield  {journal} {\bibinfo  {journal}
  {Nature Communications}\ }\textbf {\bibinfo {volume} {8}},\ \bibinfo {pages}
  {14653} (\bibinfo {year} {2017})}\BibitemShut {NoStop}%
\bibitem [{\citenamefont {Kranendonk}\ and\ \citenamefont
  {Frenkel}(1991)}]{kranendonk1991computer}%
  \BibitemOpen
  \bibfield  {author} {\bibinfo {author} {\bibfnamefont {WGT}\ \bibnamefont
  {Kranendonk}}\ and\ \bibinfo {author} {\bibfnamefont {D}~\bibnamefont
  {Frenkel}},\ }\bibfield  {title} {\enquote {\bibinfo {title} {Computer
  simulation of solid-liquid coexistence in binary hard sphere mixtures},}\
  }\href@noop {} {\bibfield  {journal} {\bibinfo  {journal} {Molecular
  physics}\ }\textbf {\bibinfo {volume} {72}},\ \bibinfo {pages} {679--697}
  (\bibinfo {year} {1991})}\BibitemShut {NoStop}%
\bibitem [{\citenamefont {Berthier}\ \emph
  {et~al.}(2016{\natexlab{a}})\citenamefont {Berthier}, \citenamefont
  {Coslovich}, \citenamefont {Ninarello},\ and\ \citenamefont
  {Ozawa}}]{BCNO2016PRL}%
  \BibitemOpen
  \bibfield  {author} {\bibinfo {author} {\bibfnamefont {Ludovic}\ \bibnamefont
  {Berthier}}, \bibinfo {author} {\bibfnamefont {Daniele}\ \bibnamefont
  {Coslovich}}, \bibinfo {author} {\bibfnamefont {Andrea}\ \bibnamefont
  {Ninarello}}, \ and\ \bibinfo {author} {\bibfnamefont {Misaki}\ \bibnamefont
  {Ozawa}},\ }\bibfield  {title} {\enquote {\bibinfo {title} {Equilibrium
  sampling of hard spheres up to the jamming density and beyond},}\ }\href
  {\doibase 10.1103/PhysRevLett.116.238002} {\bibfield  {journal} {\bibinfo
  {journal} {Phys. Rev. Lett.}\ }\textbf {\bibinfo {volume} {116}},\ \bibinfo
  {pages} {238002} (\bibinfo {year} {2016}{\natexlab{a}})}\BibitemShut
  {NoStop}%
\bibitem [{\citenamefont {Berthier}\ \emph {et~al.}(2017)\citenamefont
  {Berthier}, \citenamefont {Charbonneau}, \citenamefont {Coslovich},
  \citenamefont {Ninarello}, \citenamefont {Ozawa},\ and\ \citenamefont
  {Yaida}}]{berthier2017breaking}%
  \BibitemOpen
  \bibfield  {author} {\bibinfo {author} {\bibfnamefont {Ludovic}\ \bibnamefont
  {Berthier}}, \bibinfo {author} {\bibfnamefont {Patrick}\ \bibnamefont
  {Charbonneau}}, \bibinfo {author} {\bibfnamefont {Daniele}\ \bibnamefont
  {Coslovich}}, \bibinfo {author} {\bibfnamefont {Andrea}\ \bibnamefont
  {Ninarello}}, \bibinfo {author} {\bibfnamefont {Misaki}\ \bibnamefont
  {Ozawa}}, \ and\ \bibinfo {author} {\bibfnamefont {Sho}\ \bibnamefont
  {Yaida}},\ }\bibfield  {title} {\enquote {\bibinfo {title} {Configurational
  entropy measurements in extremely supercooled liquids that break the glass
  ceiling},}\ }\href@noop {} {\bibfield  {journal} {\bibinfo  {journal}
  {Proceedings of the National Academy of Sciences}\ }\textbf {\bibinfo
  {volume} {114}},\ \bibinfo {pages} {11356--11361} (\bibinfo {year}
  {2017})}\BibitemShut {NoStop}%
\bibitem [{\citenamefont {Rainone}\ \emph {et~al.}(2015)\citenamefont
  {Rainone}, \citenamefont {Urbani}, \citenamefont {Yoshino},\ and\
  \citenamefont {Zamponi}}]{RUYZ2015PRL}%
  \BibitemOpen
  \bibfield  {author} {\bibinfo {author} {\bibfnamefont {Corrado}\ \bibnamefont
  {Rainone}}, \bibinfo {author} {\bibfnamefont {Pierfrancesco}\ \bibnamefont
  {Urbani}}, \bibinfo {author} {\bibfnamefont {Hajime}\ \bibnamefont
  {Yoshino}}, \ and\ \bibinfo {author} {\bibfnamefont {Francesco}\ \bibnamefont
  {Zamponi}},\ }\bibfield  {title} {\enquote {\bibinfo {title} {Following the
  evolution of hard sphere glasses in infinite dimensions under external
  perturbations: Compression and shear strain},}\ }\href@noop {} {\bibfield
  {journal} {\bibinfo  {journal} {Phys. Rev. Lett.}\ }\textbf {\bibinfo
  {volume} {114}},\ \bibinfo {pages} {015701} (\bibinfo {year}
  {2015})}\BibitemShut {NoStop}%
\bibitem [{\citenamefont {Berthier}\ \emph
  {et~al.}(2016{\natexlab{b}})\citenamefont {Berthier}, \citenamefont
  {Charbonneau}, \citenamefont {Jin}, \citenamefont {Parisi}, \citenamefont
  {Seoane},\ and\ \citenamefont {Zamponi}}]{BCJPSZ2016PNAS}%
  \BibitemOpen
  \bibfield  {author} {\bibinfo {author} {\bibfnamefont {Ludovic}\ \bibnamefont
  {Berthier}}, \bibinfo {author} {\bibfnamefont {Patrick}\ \bibnamefont
  {Charbonneau}}, \bibinfo {author} {\bibfnamefont {Yuliang}\ \bibnamefont
  {Jin}}, \bibinfo {author} {\bibfnamefont {Giorgio}\ \bibnamefont {Parisi}},
  \bibinfo {author} {\bibfnamefont {Beatriz}\ \bibnamefont {Seoane}}, \ and\
  \bibinfo {author} {\bibfnamefont {Francesco}\ \bibnamefont {Zamponi}},\
  }\bibfield  {title} {\enquote {\bibinfo {title} {Growing timescales and
  lengthscales characterizing vibrations of amorphous solids},}\ }\href@noop {}
  {\bibfield  {journal} {\bibinfo  {journal} {Proc. Nat. Acad. Sci. U.S.A.}\
  }\textbf {\bibinfo {volume} {113}},\ \bibinfo {pages} {8397--8401} (\bibinfo
  {year} {2016}{\natexlab{b}})}\BibitemShut {NoStop}%
\bibitem [{\citenamefont {Jin}\ and\ \citenamefont
  {Yoshino}(2017)}]{jin2017exploring}%
  \BibitemOpen
  \bibfield  {author} {\bibinfo {author} {\bibfnamefont {Yuliang}\ \bibnamefont
  {Jin}}\ and\ \bibinfo {author} {\bibfnamefont {Hajime}\ \bibnamefont
  {Yoshino}},\ }\bibfield  {title} {\enquote {\bibinfo {title} {Exploring the
  complex free-energy landscape of the simplest glass by rheology},}\
  }\href@noop {} {\bibfield  {journal} {\bibinfo  {journal} {Nature
  Communications}\ }\textbf {\bibinfo {volume} {8}} (\bibinfo {year}
  {2017})}\BibitemShut {NoStop}%
\bibitem [{\citenamefont {Liu}\ and\ \citenamefont
  {Nagel}(1998)}]{liu1998nonlinear}%
  \BibitemOpen
  \bibfield  {author} {\bibinfo {author} {\bibfnamefont {Andrea~J}\
  \bibnamefont {Liu}}\ and\ \bibinfo {author} {\bibfnamefont {Sidney~R}\
  \bibnamefont {Nagel}},\ }\bibfield  {title} {\enquote {\bibinfo {title}
  {Nonlinear dynamics: Jamming is not just cool any more},}\ }\href@noop {}
  {\bibfield  {journal} {\bibinfo  {journal} {Nature}\ }\textbf {\bibinfo
  {volume} {396}},\ \bibinfo {pages} {21} (\bibinfo {year} {1998})}\BibitemShut
  {NoStop}%
\bibitem [{\citenamefont {Trappe}\ \emph {et~al.}(2001)\citenamefont {Trappe},
  \citenamefont {Prasad}, \citenamefont {Cipelletti}, \citenamefont {Segre},\
  and\ \citenamefont {Weitz}}]{trappe2001jamming}%
  \BibitemOpen
  \bibfield  {author} {\bibinfo {author} {\bibfnamefont {Veronique}\
  \bibnamefont {Trappe}}, \bibinfo {author} {\bibfnamefont {V}~\bibnamefont
  {Prasad}}, \bibinfo {author} {\bibfnamefont {Luca}\ \bibnamefont
  {Cipelletti}}, \bibinfo {author} {\bibfnamefont {PN}~\bibnamefont {Segre}}, \
  and\ \bibinfo {author} {\bibfnamefont {David~A}\ \bibnamefont {Weitz}},\
  }\bibfield  {title} {\enquote {\bibinfo {title} {Jamming phase diagram for
  attractive particles},}\ }\href@noop {} {\bibfield  {journal} {\bibinfo
  {journal} {Nature}\ }\textbf {\bibinfo {volume} {411}},\ \bibinfo {pages}
  {772} (\bibinfo {year} {2001})}\BibitemShut {NoStop}%
\bibitem [{\citenamefont {Ikeda}\ \emph {et~al.}(2012)\citenamefont {Ikeda},
  \citenamefont {Berthier},\ and\ \citenamefont {Sollich}}]{ikeda2012unified}%
  \BibitemOpen
  \bibfield  {author} {\bibinfo {author} {\bibfnamefont {Atsushi}\ \bibnamefont
  {Ikeda}}, \bibinfo {author} {\bibfnamefont {Ludovic}\ \bibnamefont
  {Berthier}}, \ and\ \bibinfo {author} {\bibfnamefont {Peter}\ \bibnamefont
  {Sollich}},\ }\bibfield  {title} {\enquote {\bibinfo {title} {Unified study
  of glass and jamming rheology in soft particle systems},}\ }\href@noop {}
  {\bibfield  {journal} {\bibinfo  {journal} {Physical review letters}\
  }\textbf {\bibinfo {volume} {109}},\ \bibinfo {pages} {018301} (\bibinfo
  {year} {2012})}\BibitemShut {NoStop}%
\bibitem [{\citenamefont {Biroli}\ and\ \citenamefont
  {Urbani}(2017)}]{biroli2017liu}%
  \BibitemOpen
  \bibfield  {author} {\bibinfo {author} {\bibfnamefont {Giulio}\ \bibnamefont
  {Biroli}}\ and\ \bibinfo {author} {\bibfnamefont {Pierfrancesco}\
  \bibnamefont {Urbani}},\ }\bibfield  {title} {\enquote {\bibinfo {title}
  {Liu-nagel phase diagrams in infinite dimension},}\ }\href@noop {} {\bibfield
   {journal} {\bibinfo  {journal} {arXiv preprint arXiv:1704.04649}\ }
  (\bibinfo {year} {2017})}\BibitemShut {NoStop}%
\bibitem [{\citenamefont {Urbani}\ and\ \citenamefont {Zamponi}(2017)}]{UZ17}%
  \BibitemOpen
  \bibfield  {author} {\bibinfo {author} {\bibfnamefont {Pierfrancesco}\
  \bibnamefont {Urbani}}\ and\ \bibinfo {author} {\bibfnamefont {Francesco}\
  \bibnamefont {Zamponi}},\ }\bibfield  {title} {\enquote {\bibinfo {title}
  {Shear yielding and shear jamming of dense hard sphere glasses},}\
  }\href@noop {} {\bibfield  {journal} {\bibinfo  {journal} {Physical review
  letters}\ }\textbf {\bibinfo {volume} {118}},\ \bibinfo {pages} {038001}
  (\bibinfo {year} {2017})}\BibitemShut {NoStop}%
\bibitem [{\citenamefont {Candelier}\ and\ \citenamefont
  {Dauchot}(2009)}]{candelier2009creep}%
  \BibitemOpen
  \bibfield  {author} {\bibinfo {author} {\bibfnamefont {R}~\bibnamefont
  {Candelier}}\ and\ \bibinfo {author} {\bibfnamefont {Olivier}\ \bibnamefont
  {Dauchot}},\ }\bibfield  {title} {\enquote {\bibinfo {title} {Creep motion of
  an intruder within a granular glass close to jamming},}\ }\href@noop {}
  {\bibfield  {journal} {\bibinfo  {journal} {Physical review letters}\
  }\textbf {\bibinfo {volume} {103}},\ \bibinfo {pages} {128001} (\bibinfo
  {year} {2009})}\BibitemShut {NoStop}%
\bibitem [{\citenamefont {Seguin}\ and\ \citenamefont
  {Dauchot}(2016)}]{seguin2016experimental}%
  \BibitemOpen
  \bibfield  {author} {\bibinfo {author} {\bibfnamefont {Antoine}\ \bibnamefont
  {Seguin}}\ and\ \bibinfo {author} {\bibfnamefont {Olivier}\ \bibnamefont
  {Dauchot}},\ }\bibfield  {title} {\enquote {\bibinfo {title} {Experimental
  evidences of the gardner phase in a granular glass},}\ }\href@noop {}
  {\bibfield  {journal} {\bibinfo  {journal} {Phys. Rev. Lett.}\ }\textbf
  {\bibinfo {volume} {117}},\ \bibinfo {pages} {228001} (\bibinfo {year}
  {2016})}\BibitemShut {NoStop}%
\bibitem [{\citenamefont {Falk}\ and\ \citenamefont {Langer}(1998)}]{FL98}%
  \BibitemOpen
  \bibfield  {author} {\bibinfo {author} {\bibfnamefont {Michael~L}\
  \bibnamefont {Falk}}\ and\ \bibinfo {author} {\bibfnamefont {James~S}\
  \bibnamefont {Langer}},\ }\bibfield  {title} {\enquote {\bibinfo {title}
  {Dynamics of viscoplastic deformation in amorphous solids},}\ }\href@noop {}
  {\bibfield  {journal} {\bibinfo  {journal} {Physical Review E}\ }\textbf
  {\bibinfo {volume} {57}},\ \bibinfo {pages} {7192} (\bibinfo {year}
  {1998})}\BibitemShut {NoStop}%
\bibitem [{\citenamefont {Hentschel}\ \emph {et~al.}(2011)\citenamefont
  {Hentschel}, \citenamefont {Karmakar}, \citenamefont {Lerner},\ and\
  \citenamefont {Procaccia}}]{HKLP11}%
  \BibitemOpen
  \bibfield  {author} {\bibinfo {author} {\bibfnamefont {HGE}\ \bibnamefont
  {Hentschel}}, \bibinfo {author} {\bibfnamefont {Smarajit}\ \bibnamefont
  {Karmakar}}, \bibinfo {author} {\bibfnamefont {Edan}\ \bibnamefont {Lerner}},
  \ and\ \bibinfo {author} {\bibfnamefont {Itamar}\ \bibnamefont {Procaccia}},\
  }\bibfield  {title} {\enquote {\bibinfo {title} {Do athermal amorphous solids
  exist?}}\ }\href@noop {} {\bibfield  {journal} {\bibinfo  {journal} {Physical
  Review E}\ }\textbf {\bibinfo {volume} {83}},\ \bibinfo {pages} {061101}
  (\bibinfo {year} {2011})}\BibitemShut {NoStop}%
\bibitem [{\citenamefont {Lin}\ \emph {et~al.}(2015)\citenamefont {Lin},
  \citenamefont {Gueudr{\'e}}, \citenamefont {Rosso},\ and\ \citenamefont
  {Wyart}}]{LW15}%
  \BibitemOpen
  \bibfield  {author} {\bibinfo {author} {\bibfnamefont {Jie}\ \bibnamefont
  {Lin}}, \bibinfo {author} {\bibfnamefont {Thomas}\ \bibnamefont
  {Gueudr{\'e}}}, \bibinfo {author} {\bibfnamefont {Alberto}\ \bibnamefont
  {Rosso}}, \ and\ \bibinfo {author} {\bibfnamefont {Matthieu}\ \bibnamefont
  {Wyart}},\ }\bibfield  {title} {\enquote {\bibinfo {title} {Criticality in
  the approach to failure in amorphous solids},}\ }\href@noop {} {\bibfield
  {journal} {\bibinfo  {journal} {Physical review letters}\ }\textbf {\bibinfo
  {volume} {115}},\ \bibinfo {pages} {168001} (\bibinfo {year}
  {2015})}\BibitemShut {NoStop}%
\bibitem [{\citenamefont {Franz}\ and\ \citenamefont
  {Spigler}(2016)}]{franz2016mean}%
  \BibitemOpen
  \bibfield  {author} {\bibinfo {author} {\bibfnamefont {Silvio}\ \bibnamefont
  {Franz}}\ and\ \bibinfo {author} {\bibfnamefont {Stefano}\ \bibnamefont
  {Spigler}},\ }\bibfield  {title} {\enquote {\bibinfo {title} {Mean-field
  avalanches in jammed spheres},}\ }\href@noop {} {\bibfield  {journal}
  {\bibinfo  {journal} {arXiv preprint arXiv:1608.01265}\ } (\bibinfo {year}
  {2016})}\BibitemShut {NoStop}%
\bibitem [{\citenamefont {Biroli}\ and\ \citenamefont
  {Urbani}(2016)}]{BU2016NP}%
  \BibitemOpen
  \bibfield  {author} {\bibinfo {author} {\bibfnamefont {Giulio}\ \bibnamefont
  {Biroli}}\ and\ \bibinfo {author} {\bibfnamefont {Pierfrancesco}\
  \bibnamefont {Urbani}},\ }\bibfield  {title} {\enquote {\bibinfo {title}
  {Breakdown of elasticity in amorphous solids},}\ }\href@noop {} {\bibfield
  {journal} {\bibinfo  {journal} {Nat. Phys.}\ }\textbf {\bibinfo {volume}
  {12}},\ \bibinfo {pages} {1130--1133} (\bibinfo {year} {2016})}\BibitemShut
  {NoStop}%
\bibitem [{\citenamefont {Yoshino}\ and\ \citenamefont
  {Zamponi}(2014)}]{YZ2014PRE}%
  \BibitemOpen
  \bibfield  {author} {\bibinfo {author} {\bibfnamefont {Hajime}\ \bibnamefont
  {Yoshino}}\ and\ \bibinfo {author} {\bibfnamefont {Francesco}\ \bibnamefont
  {Zamponi}},\ }\bibfield  {title} {\enquote {\bibinfo {title} {Shear modulus
  of glasses: Results from the full replica-symmetry-breaking solution},}\
  }\href@noop {} {\bibfield  {journal} {\bibinfo  {journal} {Phys. Rev. E}\
  }\textbf {\bibinfo {volume} {90}},\ \bibinfo {pages} {022302} (\bibinfo
  {year} {2014})}\BibitemShut {NoStop}%
\bibitem [{\citenamefont {Scalliet}\ \emph {et~al.}(2017)\citenamefont
  {Scalliet}, \citenamefont {Berthier},\ and\ \citenamefont
  {Zamponi}}]{PhysRevLett.119.205501}%
  \BibitemOpen
  \bibfield  {author} {\bibinfo {author} {\bibfnamefont {Camille}\ \bibnamefont
  {Scalliet}}, \bibinfo {author} {\bibfnamefont {Ludovic}\ \bibnamefont
  {Berthier}}, \ and\ \bibinfo {author} {\bibfnamefont {Francesco}\
  \bibnamefont {Zamponi}},\ }\bibfield  {title} {\enquote {\bibinfo {title}
  {Absence of marginal stability in a structural glass},}\ }\href@noop {}
  {\bibfield  {journal} {\bibinfo  {journal} {Phys. Rev. Lett.}\ }\textbf
  {\bibinfo {volume} {119}},\ \bibinfo {pages} {205501} (\bibinfo {year}
  {2017})}\BibitemShut {NoStop}%
\bibitem [{\citenamefont {Hicks}\ \emph {et~al.}(2018)\citenamefont {Hicks},
  \citenamefont {Wheatley}, \citenamefont {Godfrey},\ and\ \citenamefont
  {Moore}}]{hicks2017gardner}%
  \BibitemOpen
  \bibfield  {author} {\bibinfo {author} {\bibfnamefont {CL}~\bibnamefont
  {Hicks}}, \bibinfo {author} {\bibfnamefont {MJ}~\bibnamefont {Wheatley}},
  \bibinfo {author} {\bibfnamefont {MJ}~\bibnamefont {Godfrey}}, \ and\
  \bibinfo {author} {\bibfnamefont {MA}~\bibnamefont {Moore}},\ }\bibfield
  {title} {\enquote {\bibinfo {title} {Gardner transition in physical
  dimensions},}\ }\href@noop {} {\bibfield  {journal} {\bibinfo  {journal}
  {Physical review letters}\ }\textbf {\bibinfo {volume} {120}},\ \bibinfo
  {pages} {225501} (\bibinfo {year} {2018})}\BibitemShut {NoStop}%
\bibitem [{\citenamefont {Fullerton}\ and\ \citenamefont
  {Berthier}(2017)}]{fullerton2017density}%
  \BibitemOpen
  \bibfield  {author} {\bibinfo {author} {\bibfnamefont {Christopher~J}\
  \bibnamefont {Fullerton}}\ and\ \bibinfo {author} {\bibfnamefont {Ludovic}\
  \bibnamefont {Berthier}},\ }\bibfield  {title} {\enquote {\bibinfo {title}
  {Density controls the kinetic stability of ultrastable glasses},}\
  }\href@noop {} {\bibfield  {journal} {\bibinfo  {journal} {EPL (Europhysics
  Letters)}\ }\textbf {\bibinfo {volume} {119}},\ \bibinfo {pages} {36003}
  (\bibinfo {year} {2017})}\BibitemShut {NoStop}%
\bibitem [{\citenamefont {Singh}\ \emph {et~al.}(2013)\citenamefont {Singh},
  \citenamefont {Ediger},\ and\ \citenamefont
  {De~Pablo}}]{singh2013ultrastable}%
  \BibitemOpen
  \bibfield  {author} {\bibinfo {author} {\bibfnamefont {Sadanand}\
  \bibnamefont {Singh}}, \bibinfo {author} {\bibfnamefont {MD}~\bibnamefont
  {Ediger}}, \ and\ \bibinfo {author} {\bibfnamefont {Juan~J}\ \bibnamefont
  {De~Pablo}},\ }\bibfield  {title} {\enquote {\bibinfo {title} {Ultrastable
  glasses from in silico vapour deposition},}\ }\href@noop {} {\bibfield
  {journal} {\bibinfo  {journal} {Nature materials}\ }\textbf {\bibinfo
  {volume} {12}},\ \bibinfo {pages} {139--144} (\bibinfo {year}
  {2013})}\BibitemShut {NoStop}%
\bibitem [{\citenamefont {Karmakar}\ \emph {et~al.}(2010)\citenamefont
  {Karmakar}, \citenamefont {Lerner}, \citenamefont {Procaccia},\ and\
  \citenamefont {Zylberg}}]{karmakar2010statisticalB}%
  \BibitemOpen
  \bibfield  {author} {\bibinfo {author} {\bibfnamefont {Smarajit}\
  \bibnamefont {Karmakar}}, \bibinfo {author} {\bibfnamefont {Edan}\
  \bibnamefont {Lerner}}, \bibinfo {author} {\bibfnamefont {Itamar}\
  \bibnamefont {Procaccia}}, \ and\ \bibinfo {author} {\bibfnamefont {Jacques}\
  \bibnamefont {Zylberg}},\ }\bibfield  {title} {\enquote {\bibinfo {title}
  {Statistical physics of elastoplastic steady states in amorphous solids:
  Finite temperatures and strain rates},}\ }\href@noop {} {\bibfield  {journal}
  {\bibinfo  {journal} {Phys. Rev. E}\ }\textbf {\bibinfo {volume} {82}},\
  \bibinfo {pages} {031301} (\bibinfo {year} {2010})}\BibitemShut {NoStop}%
\bibitem [{\citenamefont {Nagata}\ \emph {et~al.}(1979)\citenamefont {Nagata},
  \citenamefont {Keesom},\ and\ \citenamefont {Harrison}}]{nagata1979low}%
  \BibitemOpen
  \bibfield  {author} {\bibinfo {author} {\bibfnamefont {Shoichi}\ \bibnamefont
  {Nagata}}, \bibinfo {author} {\bibfnamefont {PH}~\bibnamefont {Keesom}}, \
  and\ \bibinfo {author} {\bibfnamefont {HR}~\bibnamefont {Harrison}},\
  }\bibfield  {title} {\enquote {\bibinfo {title} {Low-dc-field susceptibility
  of cu mn spin glass},}\ }\href@noop {} {\bibfield  {journal} {\bibinfo
  {journal} {Physical Review B}\ }\textbf {\bibinfo {volume} {19}},\ \bibinfo
  {pages} {1633} (\bibinfo {year} {1979})}\BibitemShut {NoStop}%
\bibitem [{\citenamefont {Peyneau}\ and\ \citenamefont
  {Roux}(2008)}]{peyneau2008frictionless}%
  \BibitemOpen
  \bibfield  {author} {\bibinfo {author} {\bibfnamefont {Pierre-Emmanuel}\
  \bibnamefont {Peyneau}}\ and\ \bibinfo {author} {\bibfnamefont
  {Jean-No{\"e}l}\ \bibnamefont {Roux}},\ }\bibfield  {title} {\enquote
  {\bibinfo {title} {Frictionless bead packs have macroscopic friction, but no
  dilatancy},}\ }\href@noop {} {\bibfield  {journal} {\bibinfo  {journal}
  {Physical review E}\ }\textbf {\bibinfo {volume} {78}},\ \bibinfo {pages}
  {011307} (\bibinfo {year} {2008})}\BibitemShut {NoStop}%
\bibitem [{\citenamefont {Maxwell}(1864)}]{maxwell1864calculation}%
  \BibitemOpen
  \bibfield  {author} {\bibinfo {author} {\bibfnamefont {J~Clerk}\ \bibnamefont
  {Maxwell}},\ }\bibfield  {title} {\enquote {\bibinfo {title} {L. on the
  calculation of the equilibrium and stiffness of frames},}\ }\href@noop {}
  {\bibfield  {journal} {\bibinfo  {journal} {The London, Edinburgh, and Dublin
  Philosophical Magazine and Journal of Science}\ }\textbf {\bibinfo {volume}
  {27}},\ \bibinfo {pages} {294--299} (\bibinfo {year} {1864})}\BibitemShut
  {NoStop}%
\bibitem [{\citenamefont {Baity-Jesi}\ \emph {et~al.}(2017)\citenamefont
  {Baity-Jesi}, \citenamefont {Goodrich}, \citenamefont {Liu}, \citenamefont
  {Nagel},\ and\ \citenamefont {Sethna}}]{baity2017emergent}%
  \BibitemOpen
  \bibfield  {author} {\bibinfo {author} {\bibfnamefont {Marco}\ \bibnamefont
  {Baity-Jesi}}, \bibinfo {author} {\bibfnamefont {Carl~P}\ \bibnamefont
  {Goodrich}}, \bibinfo {author} {\bibfnamefont {Andrea~J}\ \bibnamefont
  {Liu}}, \bibinfo {author} {\bibfnamefont {Sidney~R}\ \bibnamefont {Nagel}}, \
  and\ \bibinfo {author} {\bibfnamefont {James~P}\ \bibnamefont {Sethna}},\
  }\bibfield  {title} {\enquote {\bibinfo {title} {Emergent so (3) symmetry of
  the frictionless shear jamming transition},}\ }\href@noop {} {\bibfield
  {journal} {\bibinfo  {journal} {Journal of Statistical Physics}\ }\textbf
  {\bibinfo {volume} {167}},\ \bibinfo {pages} {735--748} (\bibinfo {year}
  {2017})}\BibitemShut {NoStop}%
\bibitem [{\citenamefont {Urbani}\ and\ \citenamefont {Biroli}(2015)}]{BU15}%
  \BibitemOpen
  \bibfield  {author} {\bibinfo {author} {\bibfnamefont {Pierfrancesco}\
  \bibnamefont {Urbani}}\ and\ \bibinfo {author} {\bibfnamefont {Giulio}\
  \bibnamefont {Biroli}},\ }\bibfield  {title} {\enquote {\bibinfo {title}
  {Gardner transition in finite dimensions},}\ }\href@noop {} {\bibfield
  {journal} {\bibinfo  {journal} {Physical Review B}\ }\textbf {\bibinfo
  {volume} {91}},\ \bibinfo {pages} {100202} (\bibinfo {year}
  {2015})}\BibitemShut {NoStop}%
\bibitem [{\citenamefont {Charbonneau}\ and\ \citenamefont
  {Yaida}(2017)}]{charbonneau2017nontrivial}%
  \BibitemOpen
  \bibfield  {author} {\bibinfo {author} {\bibfnamefont {Patrick}\ \bibnamefont
  {Charbonneau}}\ and\ \bibinfo {author} {\bibfnamefont {Sho}\ \bibnamefont
  {Yaida}},\ }\bibfield  {title} {\enquote {\bibinfo {title} {Nontrivial
  critical fixed point for replica-symmetry-breaking transitions},}\
  }\href@noop {} {\bibfield  {journal} {\bibinfo  {journal} {Physical review
  letters}\ }\textbf {\bibinfo {volume} {118}},\ \bibinfo {pages} {215701}
  (\bibinfo {year} {2017})}\BibitemShut {NoStop}%
\bibitem [{\citenamefont {Lubchenko}\ and\ \citenamefont
  {Wolynes}(2017)}]{lubchenko2017aging}%
  \BibitemOpen
  \bibfield  {author} {\bibinfo {author} {\bibfnamefont {Vassiliy}\
  \bibnamefont {Lubchenko}}\ and\ \bibinfo {author} {\bibfnamefont {Peter~G}\
  \bibnamefont {Wolynes}},\ }\bibfield  {title} {\enquote {\bibinfo {title}
  {Aging, jamming, and the limits of stability of amorphous solids},}\
  }\href@noop {} {\bibfield  {journal} {\bibinfo  {journal} {The Journal of
  Physical Chemistry B}\ } (\bibinfo {year} {2017})}\BibitemShut {NoStop}%
\bibitem [{\citenamefont {Rainone}\ and\ \citenamefont
  {Urbani}(2016)}]{CP2016}%
  \BibitemOpen
  \bibfield  {author} {\bibinfo {author} {\bibfnamefont {Corrado}\ \bibnamefont
  {Rainone}}\ and\ \bibinfo {author} {\bibfnamefont {Pierfrancesco}\
  \bibnamefont {Urbani}},\ }\bibfield  {title} {\enquote {\bibinfo {title}
  {Following the evolution of glassy states under external perturbations: the
  full replica symmetry breaking solution},}\ }\href@noop {} {\bibfield
  {journal} {\bibinfo  {journal} {J. Stat. Mech. Theor. Exp.}\ }\textbf
  {\bibinfo {volume} {2016}},\ \bibinfo {pages} {053302} (\bibinfo {year}
  {2016})}\BibitemShut {NoStop}%
\bibitem [{\citenamefont {Jaiswal}\ \emph {et~al.}(2016)\citenamefont
  {Jaiswal}, \citenamefont {Procaccia}, \citenamefont {Rainone},\ and\
  \citenamefont {Singh}}]{JPRS16}%
  \BibitemOpen
  \bibfield  {author} {\bibinfo {author} {\bibfnamefont {Prabhat~K}\
  \bibnamefont {Jaiswal}}, \bibinfo {author} {\bibfnamefont {Itamar}\
  \bibnamefont {Procaccia}}, \bibinfo {author} {\bibfnamefont {Corrado}\
  \bibnamefont {Rainone}}, \ and\ \bibinfo {author} {\bibfnamefont {Murari}\
  \bibnamefont {Singh}},\ }\bibfield  {title} {\enquote {\bibinfo {title}
  {Mechanical yield in amorphous solids: A first-order phase transition},}\
  }\href@noop {} {\bibfield  {journal} {\bibinfo  {journal} {Physical review
  letters}\ }\textbf {\bibinfo {volume} {116}},\ \bibinfo {pages} {085501}
  (\bibinfo {year} {2016})}\BibitemShut {NoStop}%
\bibitem [{\citenamefont {Parisi}\ \emph {et~al.}(2017)\citenamefont {Parisi},
  \citenamefont {Procaccia}, \citenamefont {Rainone},\ and\ \citenamefont
  {Singh}}]{PPRS17}%
  \BibitemOpen
  \bibfield  {author} {\bibinfo {author} {\bibfnamefont {Giorgio}\ \bibnamefont
  {Parisi}}, \bibinfo {author} {\bibfnamefont {Itamar}\ \bibnamefont
  {Procaccia}}, \bibinfo {author} {\bibfnamefont {Corrado}\ \bibnamefont
  {Rainone}}, \ and\ \bibinfo {author} {\bibfnamefont {Murari}\ \bibnamefont
  {Singh}},\ }\bibfield  {title} {\enquote {\bibinfo {title} {Shear bands as
  manifestation of a criticality in yielding amorphous solids},}\ }\href@noop
  {} {\bibfield  {journal} {\bibinfo  {journal} {Proceedings of the National
  Academy of Sciences}\ }\textbf {\bibinfo {volume} {114}},\ \bibinfo {pages}
  {5577--5582} (\bibinfo {year} {2017})}\BibitemShut {NoStop}%
\bibitem [{\citenamefont {Wisitsorasak}\ and\ \citenamefont
  {Wolynes}(2012)}]{wisitsorasak2012strength}%
  \BibitemOpen
  \bibfield  {author} {\bibinfo {author} {\bibfnamefont {Apiwat}\ \bibnamefont
  {Wisitsorasak}}\ and\ \bibinfo {author} {\bibfnamefont {Peter~G}\
  \bibnamefont {Wolynes}},\ }\bibfield  {title} {\enquote {\bibinfo {title} {On
  the strength of glasses},}\ }\href@noop {} {\bibfield  {journal} {\bibinfo
  {journal} {Proceedings of the National Academy of Sciences}\ }\textbf
  {\bibinfo {volume} {109}},\ \bibinfo {pages} {16068--16072} (\bibinfo {year}
  {2012})}\BibitemShut {NoStop}%
\bibitem [{\citenamefont {Manning}\ and\ \citenamefont {Liu}(2011)}]{ML11}%
  \BibitemOpen
  \bibfield  {author} {\bibinfo {author} {\bibfnamefont {M~Lisa}\ \bibnamefont
  {Manning}}\ and\ \bibinfo {author} {\bibfnamefont {Andrea~J}\ \bibnamefont
  {Liu}},\ }\bibfield  {title} {\enquote {\bibinfo {title} {Vibrational modes
  identify soft spots in a sheared disordered packing},}\ }\href@noop {}
  {\bibfield  {journal} {\bibinfo  {journal} {Physical Review Letters}\
  }\textbf {\bibinfo {volume} {107}},\ \bibinfo {pages} {108302} (\bibinfo
  {year} {2011})}\BibitemShut {NoStop}%
\bibitem [{\citenamefont {Mizuno}\ \emph {et~al.}(2017)\citenamefont {Mizuno},
  \citenamefont {Shiba},\ and\ \citenamefont {Ikeda}}]{Mizuno14112017}%
  \BibitemOpen
  \bibfield  {author} {\bibinfo {author} {\bibfnamefont {Hideyuki}\
  \bibnamefont {Mizuno}}, \bibinfo {author} {\bibfnamefont {Hayato}\
  \bibnamefont {Shiba}}, \ and\ \bibinfo {author} {\bibfnamefont {Atsushi}\
  \bibnamefont {Ikeda}},\ }\bibfield  {title} {\enquote {\bibinfo {title}
  {Continuum limit of the vibrational properties of amorphous solids},}\
  }\href@noop {} {\bibfield  {journal} {\bibinfo  {journal} {Proceedings of the
  National Academy of Sciences}\ }\textbf {\bibinfo {volume} {114}},\ \bibinfo
  {pages} {E9767--E9774} (\bibinfo {year} {2017})}\BibitemShut {NoStop}%
\bibitem [{\citenamefont {Lerner}\ \emph {et~al.}(2016)\citenamefont {Lerner},
  \citenamefont {D{\"u}ring},\ and\ \citenamefont
  {Bouchbinder}}]{lerner2016statistics}%
  \BibitemOpen
  \bibfield  {author} {\bibinfo {author} {\bibfnamefont {Edan}\ \bibnamefont
  {Lerner}}, \bibinfo {author} {\bibfnamefont {Gustavo}\ \bibnamefont
  {D{\"u}ring}}, \ and\ \bibinfo {author} {\bibfnamefont {Eran}\ \bibnamefont
  {Bouchbinder}},\ }\bibfield  {title} {\enquote {\bibinfo {title} {Statistics
  and properties of low-frequency vibrational modes in structural glasses},}\
  }\href@noop {} {\bibfield  {journal} {\bibinfo  {journal} {Physical review
  letters}\ }\textbf {\bibinfo {volume} {117}},\ \bibinfo {pages} {035501}
  (\bibinfo {year} {2016})}\BibitemShut {NoStop}%
\bibitem [{\citenamefont {Fiocco}\ \emph {et~al.}(2013)\citenamefont {Fiocco},
  \citenamefont {Foffi},\ and\ \citenamefont {Sastry}}]{fiocco2013oscillatory}%
  \BibitemOpen
  \bibfield  {author} {\bibinfo {author} {\bibfnamefont {Davide}\ \bibnamefont
  {Fiocco}}, \bibinfo {author} {\bibfnamefont {Giuseppe}\ \bibnamefont
  {Foffi}}, \ and\ \bibinfo {author} {\bibfnamefont {Srikanth}\ \bibnamefont
  {Sastry}},\ }\bibfield  {title} {\enquote {\bibinfo {title} {Oscillatory
  athermal quasistatic deformation of a model glass},}\ }\href@noop {}
  {\bibfield  {journal} {\bibinfo  {journal} {Physical Review E}\ }\textbf
  {\bibinfo {volume} {88}},\ \bibinfo {pages} {020301} (\bibinfo {year}
  {2013})}\BibitemShut {NoStop}%
\bibitem [{\citenamefont {Lubachevsky}\ and\ \citenamefont
  {Stillinger}(1990)}]{lubachevsky1990geometric}%
  \BibitemOpen
  \bibfield  {author} {\bibinfo {author} {\bibfnamefont {Boris~D}\ \bibnamefont
  {Lubachevsky}}\ and\ \bibinfo {author} {\bibfnamefont {Frank~H}\ \bibnamefont
  {Stillinger}},\ }\bibfield  {title} {\enquote {\bibinfo {title} {Geometric
  properties of random disk packings},}\ }\href@noop {} {\bibfield  {journal}
  {\bibinfo  {journal} {Journal of statistical Physics}\ }\textbf {\bibinfo
  {volume} {60}},\ \bibinfo {pages} {561--583} (\bibinfo {year}
  {1990})}\BibitemShut {NoStop}%
\bibitem [{\citenamefont {Lees}\ and\ \citenamefont
  {Edwards}(1972)}]{lees1972computer}%
  \BibitemOpen
  \bibfield  {author} {\bibinfo {author} {\bibfnamefont {AW}~\bibnamefont
  {Lees}}\ and\ \bibinfo {author} {\bibfnamefont {SF}~\bibnamefont {Edwards}},\
  }\bibfield  {title} {\enquote {\bibinfo {title} {The computer study of
  transport processes under extreme conditions},}\ }\href@noop {} {\bibfield
  {journal} {\bibinfo  {journal} {J. Phys. Condens. Matter}\ }\textbf {\bibinfo
  {volume} {5}},\ \bibinfo {pages} {1921} (\bibinfo {year} {1972})}\BibitemShut
  {NoStop}%
\end{thebibliography}%

\begin{acknowledgments}
We warmly thank L.~Berthier, M.~Ozawa, C.~Scalliet, M.~Wyart, A.~Altieri,  O.~Dauchot, {\color{black} K.~Miyazaki, T.~Kawasaki} for discussions.
This work was supported by KAKENHI (No. 25103005  ``Fluctuation \& Structure'' and No. 50335337) from MEXT, Japan,
{\color{black} by the Chinese Academy of Sciences  Pioneer Hundred-Talent Program (Yuliang Jin), }
by a grant from the Simons Foundation (\#454955, Francesco Zamponi), and
by ``Investissements d'Avenir" LabEx PALM (ANR-10-LABX-0039-PALM) (Pierfrancesco Urbani).
  The computations were performed using the computing facilities in Research Center for Computational Science, Okazaki, Japan and in 
  the Cybermedia center, Osaka University.\\
\end{acknowledgments}

{\color{black} Competing Interests: The authors declare that they have no competing interests.\\

Author contributions:  All authors designed research. Y.~J. wrote the code and performed the
numerical simulations and the data analysis,
in close collaboration with H.~Y. All authors contributed to the data
analysis, the theoretical interpretation of the results, and writing the manuscript. \\

Data availability: All data needed to evaluate the conclusions in the paper are present in the paper and/or the Supplementary Materials. Additional data available from authors upon request. \\

Human or animal subjects: We do not have human or animal subjects. }

\clearpage
\setcounter{figure}{0}  
\setcounter{equation}{0}  
\renewcommand\thefigure{S\arabic{figure}}
\renewcommand\theequation{S\arabic{equation}}

\centerline{\Large \bf Supplementary Materials}

\begin{figure}[h]
\centerline{\includegraphics[width=0.4\columnwidth]{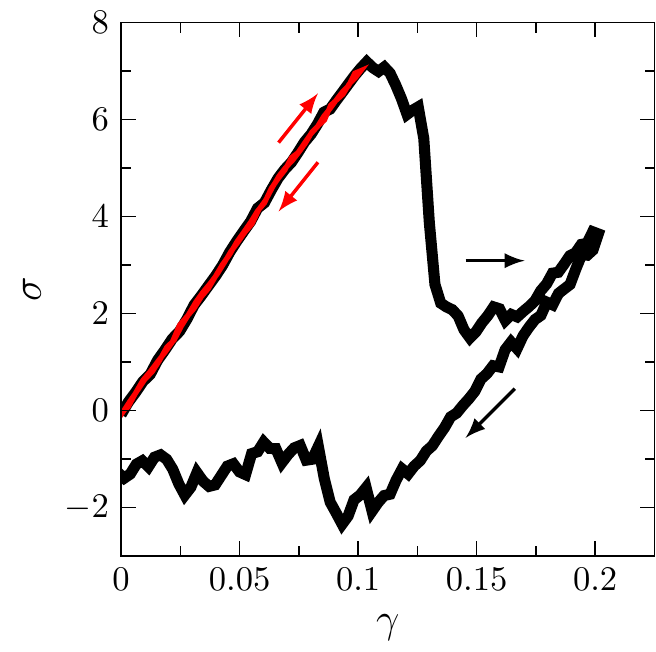}}
\caption{{\bf Singe-realization stress-strain curve for $\varphi_{\rm g} =0.655$ and $\epsilon = 0.057$.} Single-realization stress-strain curve of the same glass sample as in Fig.~1 at the fixed {\color{black} volume} strain $\epsilon = 0.057$ (or $\varphi = 0.62$). The shear strain is reversed at $\gamma = 0.1$ (red) and $\gamma = 0.2$ (black). In contrast to the case in Fig.~1A, the partially irreversible  regime is not observed. Note that according to the stability-reversibility map of Fig.~2, the Gardner line will not be crossed over in the CV-S protocol with the {\color{black} volume} strain $\epsilon = 0.057$.}
\end{figure}

\begin{figure}
\centerline{\includegraphics[width=0.7\columnwidth]{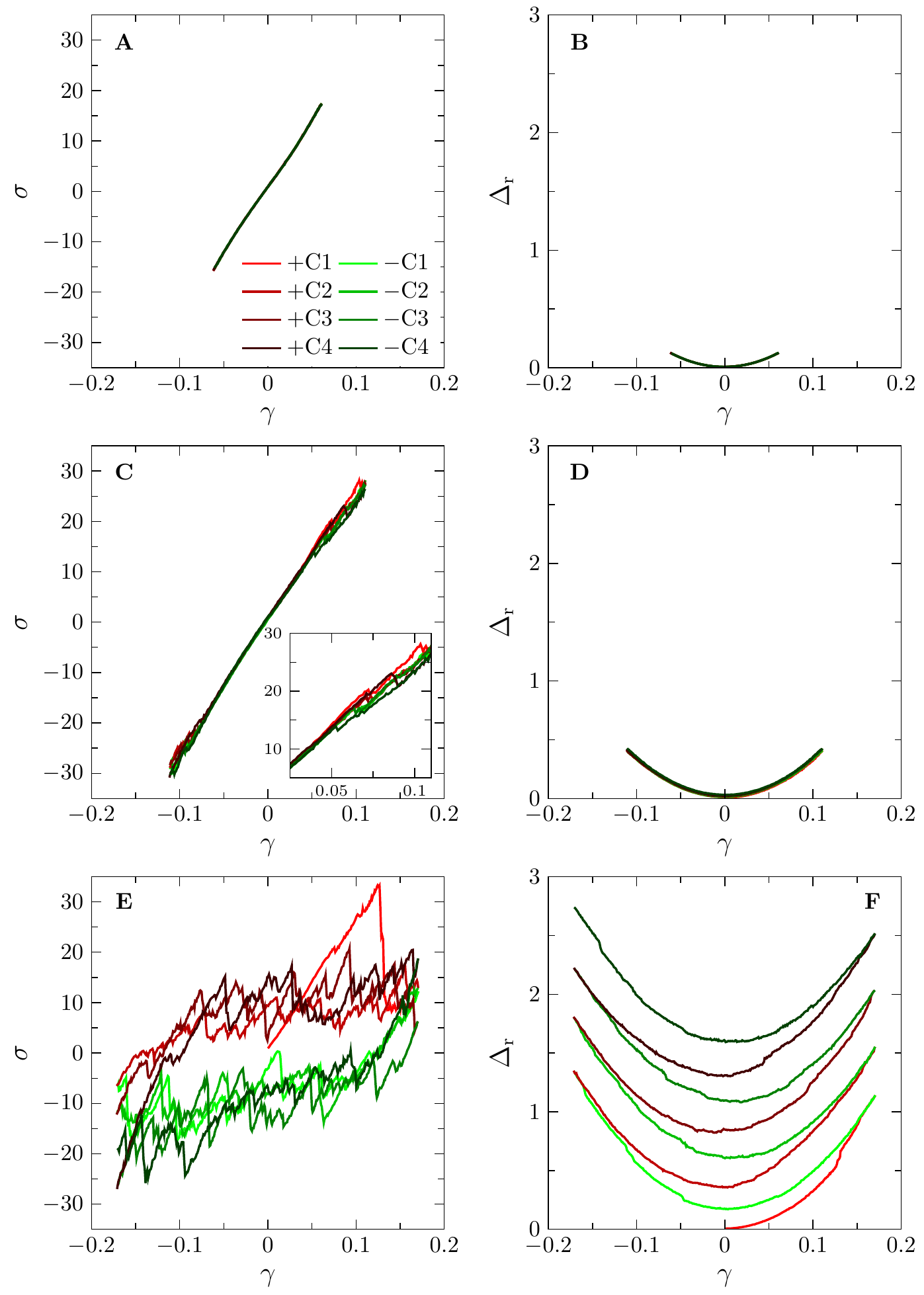}}
\caption{{\bf Multi-cycle stress-strain curves.} {\color{black} Single-realization stress-strain} curves of a single sample over four cycles of constant volume shear at $\epsilon = -0.0069$, or  $\varphi = 0.66$ (compressed from $\varphi_{\rm g} = 0.655$). The shear strain is reversed at {\bf (A)} $\gamma = \pm 0.06$,  {\bf (C)} $\gamma = \pm 0.11$, and {\bf (E)} $\gamma = \pm 0.17$. The cycle numbers (1, 2, 3, 4) and the shear directions ($+$ or $-$) are indicated.  In  {\bf (C)}, the data for $0.02 \leq \gamma \leq 0.11$ are magnified in the inset to show better the plastic events. The corresponding data of the relative mean squared displacement $\Delta_{\rm r}$ 
are shown in {\bf (B)}, {\bf (D)} and {\bf (F)}.  
While in {\bf (B)} and {\bf (D)} the system returns to the initial state,
in {\bf (F)} a diffusive behavior of $\Delta_{\rm r}$, which increases steadily at each cycle, is observed.
 }
\end{figure}

\begin{figure}
\centerline{\includegraphics[width=0.9\columnwidth]{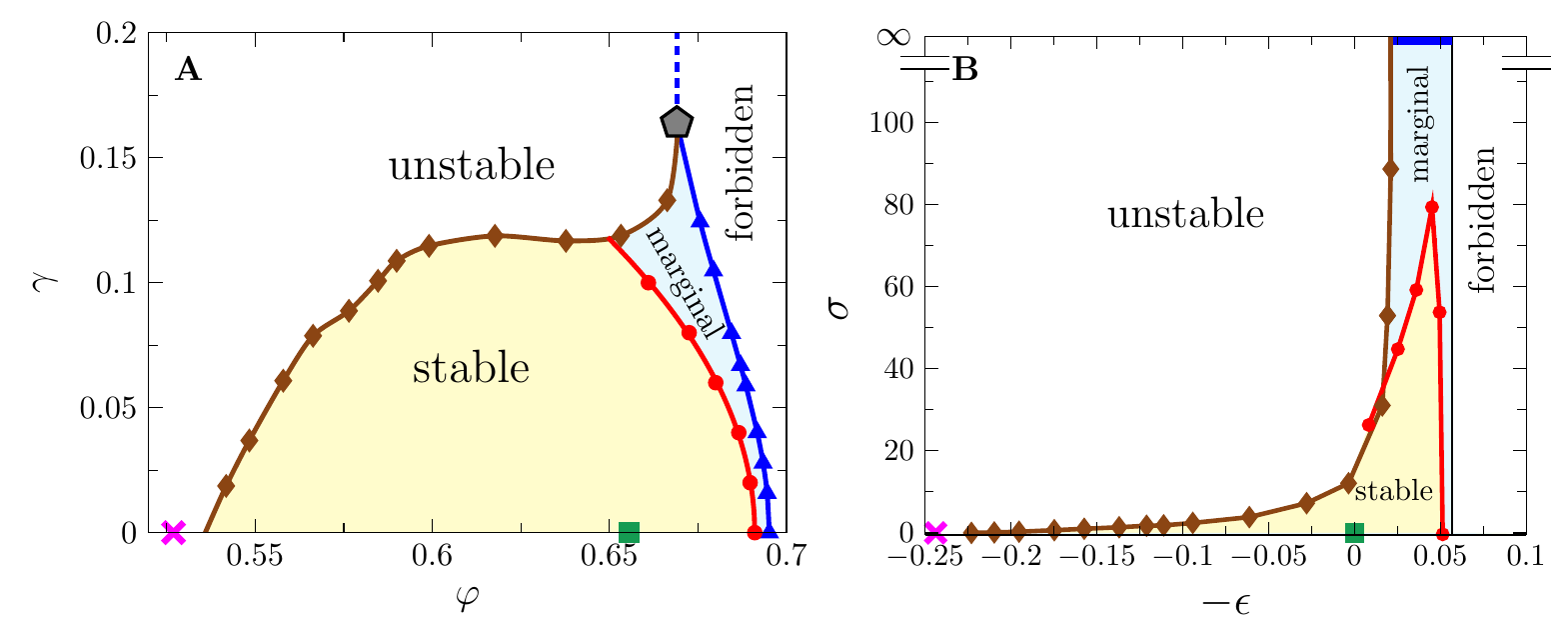}}
\caption{{\bf Other representations of the stability-reversibility map.}
Stability-reversibility map of the HS glass annealed up to  $\varphi_{\rm g} = 0.655$, 
as in Fig.~2, {\color{black} but represented in terms of {\bf (A)} volume  fraction $\varphi$ and shear strain $\gamma$, and {\bf (B)} {\color{black} volume} strain $\epsilon$ and shear stress $\sigma$. See Fig.~2 for the meaning of symbols.}
}
\end{figure}

\begin{figure}
\centerline{\includegraphics[width=0.4\columnwidth]{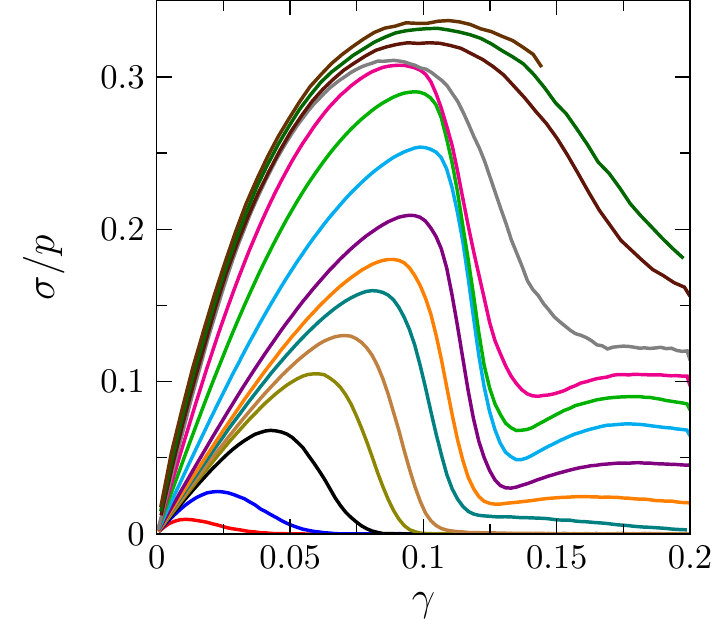}}
\caption{{\bf Rescaled stress-strain curves.} The ratio $\sigma/p$ is plotted as a function of $\gamma$ in the CP-S protocol, for  
$p =$  14.5, 15.0, 15.8, 16.5, 17, 18, 19, 21, 27, 40, 65, 160,  1000, 3000, 10000 (from bottom to top), and $\varphi_{\rm g} = 0.655$.}
\end{figure}


\begin{figure}
\centerline{\includegraphics[width=0.8\columnwidth]{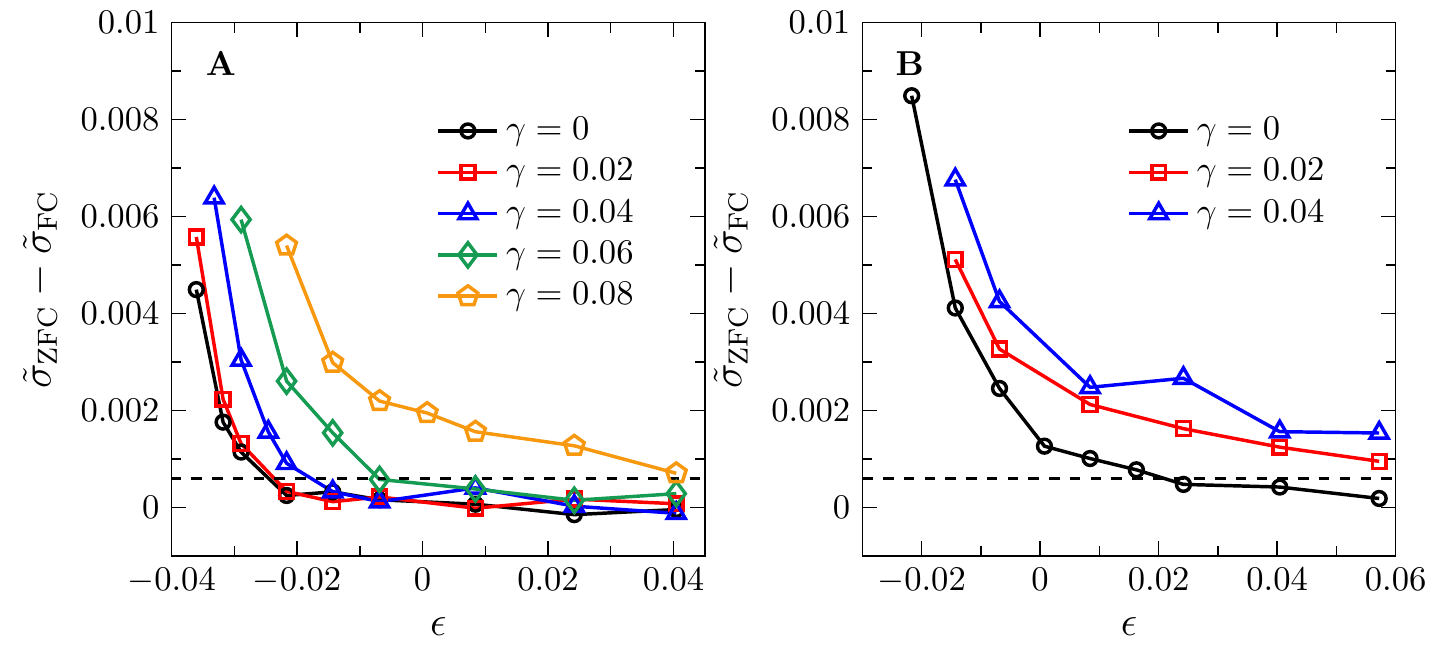}}
\caption{{\bf Determination of the Gardner threshold for other $\varphi_{\rm g}$.} The difference between ZFC and FC stresses (rescaled by $p$, $\tilde \sigma  = \sigma/p$) as a function of {\color{black} volume} strain $\epsilon$ for {\bf (A)} $\varphi_{\rm g} = 0.631$ and {\bf (B)} $\varphi_{\rm g} = 0.609$. Data for a few different $\gamma$ are plotted. The horizontal dashed lines represent the threshold value 0.0006 used to determine $\epsilon_{\rm G}$.}
\end{figure}

\begin{figure}
\centerline{\includegraphics[width=0.4\columnwidth]{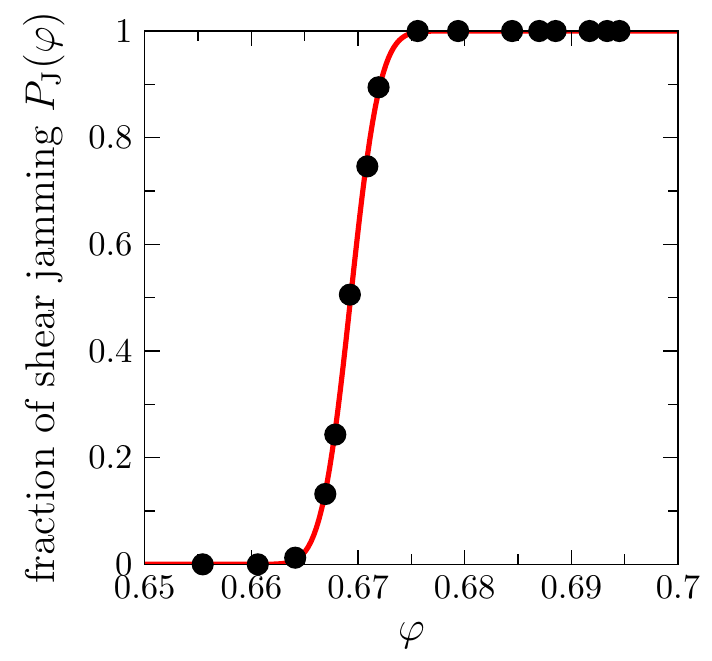}}
\caption{{\bf Determination of the yielding-jamming crossover point.} Fraction of shear jamming $P_{\rm J}(\varphi)$ as a function of $\varphi$. We use the following criteria to define shear jamming and yielding: a system jams with increasing $\gamma$ if its pressure $p$ exceeds $10^5$; otherwise, if the system can reach the maximum strain $\gamma_{\rm max} =0.2$ without jamming, then it yields. We use $P_{\rm J}(\varphi)$ to denote the fraction of shear jammed  realizations among $N_{\rm r} = 300 - 1200$ total realizations. 
The data are fitted to the error function form $P_{\rm J} (\varphi) = \frac{1}{2}+ \frac{1}{2} \erf[(\varphi - \varphi_{\rm c})/w ]$ (line), where $\varphi_{\rm c} = 0.66931(3)$ and $w = 0.0031(1)$ are fitting parameters. }
\end{figure}

\begin{figure}
\centerline{\includegraphics[width=0.8\columnwidth]{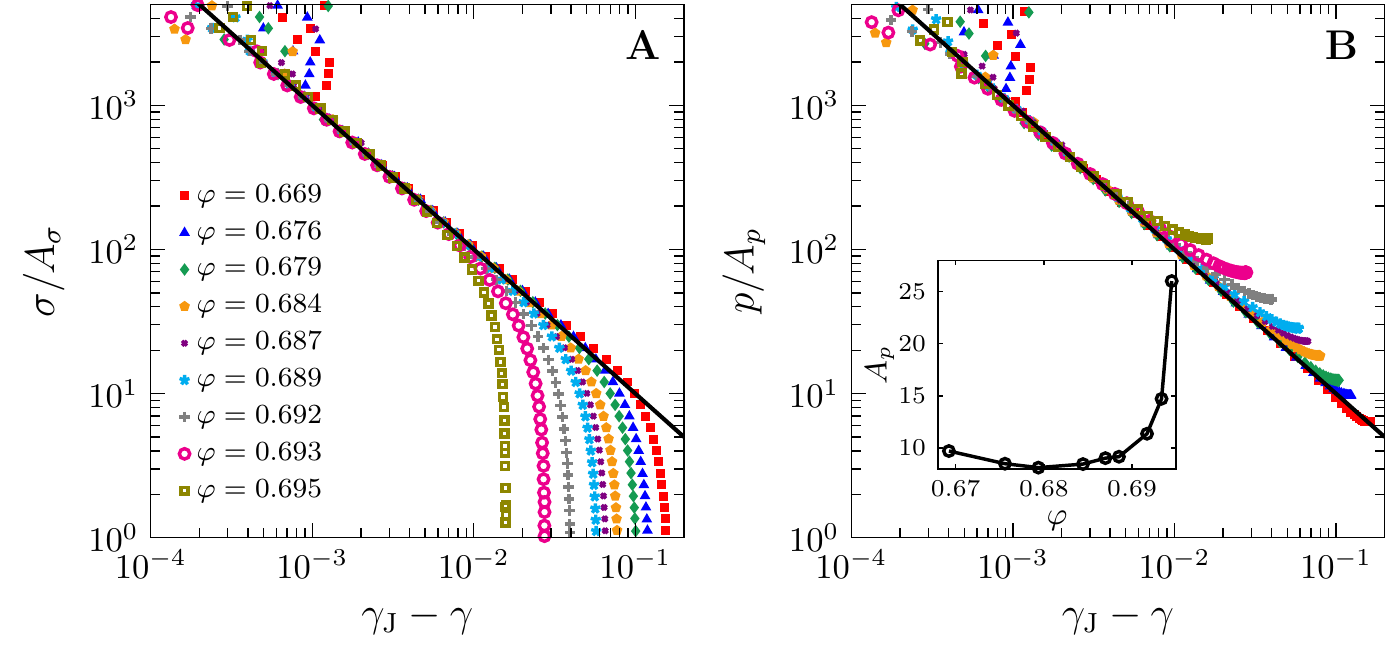}}
\caption{{\bf Free-volume scalings in shear-jamming.} The simulation data for different $\varphi$ obtained by the CV-S protocol ($\varphi_{\rm g} = 0.655$) are fitted to the free-volume scaling laws {\bf (A)} $\sigma = A_\sigma (\gamma_{\rm J} - \gamma)^{-1}$ and {\bf (B)} $p = A_p (\gamma_{\rm J} - \gamma)^{-1}$, where $A_\sigma$, $A_p$, and $\gamma_{\rm J}$ are fitting parameters. The values of $\gamma_{\rm J}$ are used to determine the shear-jamming line in the main text. We find that $A_\sigma \simeq 2.6$ is nearly independent of $\varphi$. The values of $A_p$ are plotted in the inset of {\bf (B)}. 
}
\end{figure}

\begin{figure}
\centerline{\includegraphics[width=0.4\columnwidth]{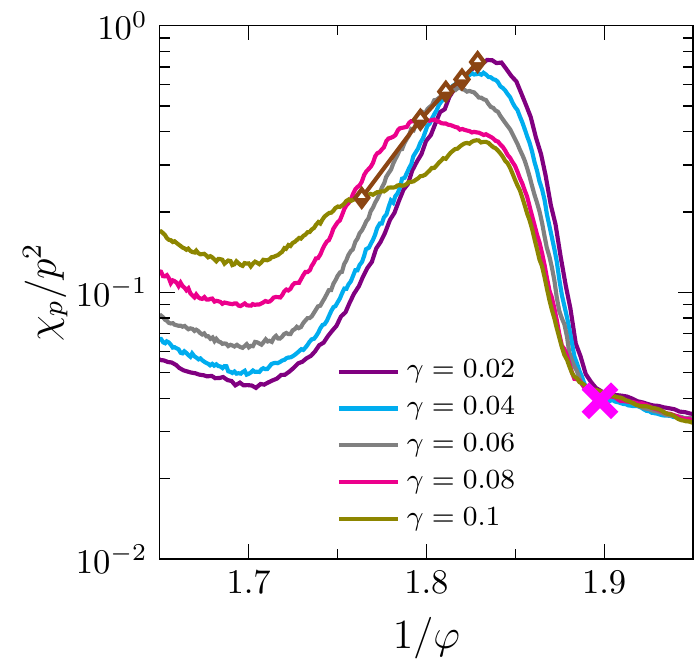}}
\caption{ {\bf Pressure susceptibility in the constant shear-compression/decompression (CS-C/D) protocol.} The pressure susceptibility $\chi_{p}=N (\langle p^2 \rangle - \langle p \rangle^2)$ (rescaled by $p^2$) as a function of $1/\varphi$ for a few different $\gamma$ in the {\color{black} CS-C/D} protocol. In contrast to $\chi_\sigma$ (Fig. 5I),  the pressure susceptibility $\chi_{p} $ has two peaks  at large $\gamma$. The first peak, caused by melting, is independent of $\gamma$, while the second one, corresponding to yielding, is at a location consistent with the peak of $\chi_\sigma$ (half filled brown diamonds). The $\gamma$-independence of the melting peak in $\chi_{p}$  further confirms that melting is independent of shear strain. For small $\gamma$, the two peaks are indistinguishable. The data suggest that the signature of melting only appears in the pressure susceptibility, but not  in the stress susceptibility.}
\end{figure}


\begin{figure}
\centerline{\includegraphics[width=\columnwidth]{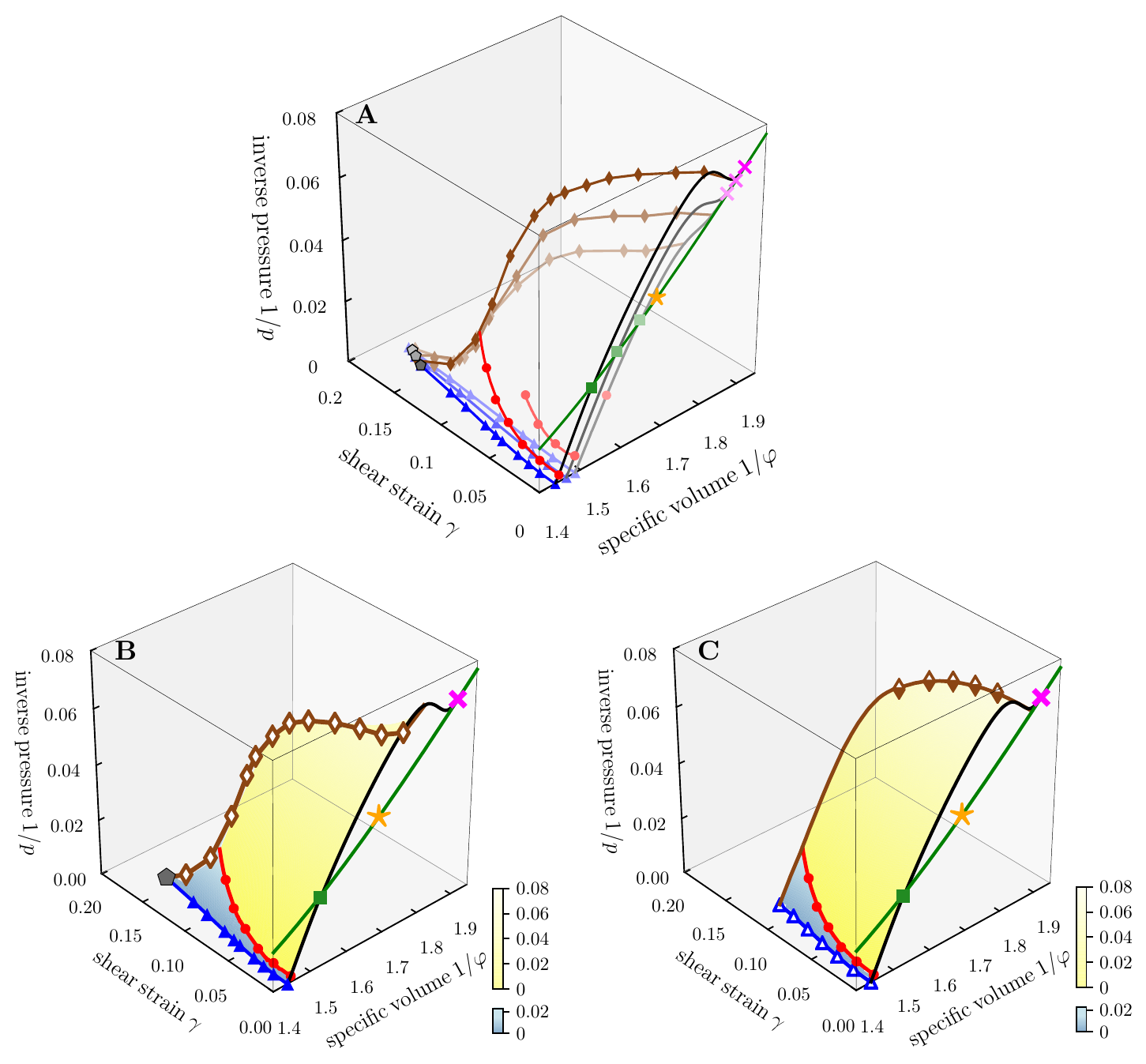}}
\caption{ {\bf Dependence of the stability-reversibility map on $\varphi_{\rm g}$ and protocols.} {\color{black} {\bf (A)} Three dimensional view of the stability-reversibility maps for  $\varphi_{\rm g} = 0.609, 0.631, 0.655$ (lighter colors represent lower $\varphi_{\rm g}$) obtained by using the CP-S protocol. {\bf (B-C)} The same plot 
for $\varphi_{\rm g} = 0.655$, obtained by using the {\bf (B)} CV-S and {\bf (C)} CS-C/D protocols.} See Fig.~2 for the meaning of the symbols.}
\end{figure}

\begin{figure}
\centerline{\includegraphics[width=0.5\columnwidth]{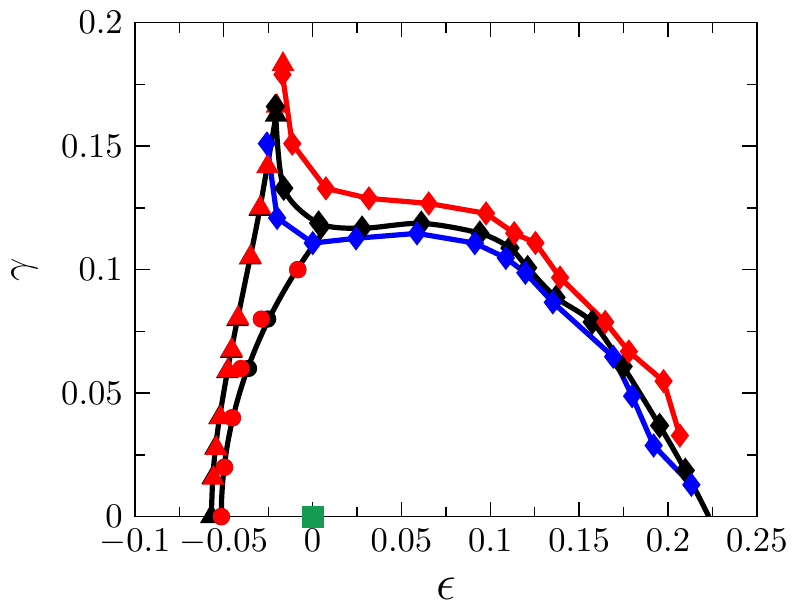}}
\caption{{\bf Dependence of the stability-reversibility map on the system size.} Stability-reversibility maps for $N = 500$ (red) and  $N=1000$ (black) systems ($\varphi_{\rm g} = 0.655$). No appreciable $N$-dependence is observed for the shear-jamming line and the Gardner line. We also plot the shear-yielding line for $N=2000$ systems (blue), showing that larger systems have lower yielding strain $\gamma_{\rm Y}$. See Fig.~2 for the meaning of the symbols. }
\end{figure}

\clearpage

{\bf Text S1. Bare and macro shear moduli.}

As discussed in the main text, two shear moduli can be defined for glasses: the bare modulus 
\beq
\mu_{\rm bare}=\lim_{N \to \infty} \lim_{\delta \gamma \to 0} \delta \sigma(\varphi_{\rm g}; \epsilon, \gamma; N)/\delta \gamma,
\eeq  
and the macroscopic modulus
\beq
\mu_{\rm macro}=\lim_{\delta \gamma \to 0} \lim_{N \to \infty}\delta \sigma(\varphi_{\rm g}; \epsilon, \gamma; N)/\delta \gamma.
\eeq
According to the mean-field theory~{\color{black} [37]}
in stable glasses $\mu_{\rm bare} = \mu_{\rm macro}$, while in marginal glasses $\mu_{\rm bare} > \mu_{\rm macro}$. In particular,  the two shear moduli have different large-$p$ scalings in the marginal phase, $\mu_{\rm macro} \sim p$ and $\mu_{\rm bare} \sim p^{\kappa}$, where $\kappa \sim 1.41574$.

In principle, we expect that the zero-field compression (ZFC) modulus $\mu_{\rm ZFC}$ and the field compression (FC) modulus $\mu_{\rm FC}$ measured in simulations have the correspondence $\mu_{\rm ZFC} \sim \mu_{\rm bare}$ and $\mu_{\rm ZFC} \sim \mu_{\rm macro}$. Ref.~{\color{black}[24]}
shows that the simulation results of three dimensional HS glasses are generally consistent with the above theoretical predictions. In the marginal phase, $\mu_{\rm ZFC}$ and $\mu_{\rm FC}$ clearly have different scalings with $p$. It was  also found that, at large $p$,  $\mu_{\rm ZFC}$ decreases with increasing $N$ or $\delta \gamma$ (note that in simulations, the modulus is measured as $\mu = \delta \sigma /\delta \gamma$, where small, but finite $\delta \gamma$ is used). This shows that the order of limits $N \to \infty$ and $\delta \gamma \to 0$ is important in the definition of shear modulus. If we fix a finite $N$, then by increasing $\delta \gamma$, $\mu_{\rm ZFC} \to \mu_{\rm FC} = \mu_{\rm macro}$. In  fact, one should only be able to detect the $\mu_{\rm ZFC}$ if $\delta \gamma < \delta \gamma_{\rm trigger}$ as discussed in the main text. In this study, we use a small enough $\delta \gamma = 0.002$, as shown in~{\color{black}[24],
to measure $\mu_{\rm ZFC}$ and  $\mu_{\rm FC}$.
  
In the measurements of the glass equation of state (G-EOS), either the constant volume-shear (CV-S) or the constant pressure-shear (CP-S) protocol corresponds to ZFC. However, we find that the curves $\sigma/p$ versus $\gamma$ collapse for large $p$ (see Fig.~S{\color{blue}4}), implying a scaling $\sigma \sim p$, as $p \to \infty$, similar to $\mu_{\rm macro} \sim p$. The result confirms that for large $\gamma$, the plasticity events are averaged out in the stress, and therefore only the macroscopic stress  $\sigma_{\rm macro}$ can be measured. This is the reason why the G-EOS itself does not encode the 
signal associated to the Gardner phase.

\end{document}